\pdfoutput=1

\documentclass[11pt]{article}
\usepackage{setspace}
\usepackage{hyperref}
\usepackage[]{ACL2023}
\usepackage{times}
\usepackage{latexsym}

\usepackage[T1]{fontenc}

\usepackage[utf8]{inputenc}

\usepackage{microtype}

\usepackage{inconsolata}
\usepackage{graphicx}
\usepackage{stfloats}
\usepackage{multirow}
\usepackage{subfigure} 
\usepackage{makecell}
\usepackage{caption}
\usepackage{subfigure}
\usepackage{arydshln}
\usepackage{booktabs}
\usepackage{multirow}
\usepackage{tablefootnote}
\usepackage{algorithm}
\usepackage{algorithmic}
\usepackage{amsmath}
\usepackage{ulem}
\usepackage{multirow}
\usepackage{arydshln}
\usepackage{booktabs}
\usepackage{newfloat}
\usepackage{listings}
\usepackage{soul}
\usepackage{float}
\usepackage{amsfonts}
\title{BERM: Training the Balanced and Extractable Representation for Matching to Improve Generalization Ability of Dense Retrieval}

 \author{
Shicheng Xu$^{1,2}$,
Liang Pang$^{1}$\thanks{\ \ Corresponding authors},
Huawei Shen$^{1,2}$,
Xueqi Cheng$^{1,2}$\footnotemark[1]\\
$^{1}$Data Intelligence System Research Center,\\
 Institute of Computing Technology, CAS\\
 $^{2}$University of Chinese Academy of Sciences \\
{\tt\ xschit@163.com \{pangliang,shenhuawei,cxq\}@ict.ac.cn}}

\begin{document}
\maketitle
\begin{abstract}

Dense retrieval has shown promise in the first-stage retrieval process when trained on in-domain labeled datasets. However, previous studies have found that dense retrieval is hard to generalize to unseen domains due to its weak modeling of domain-invariant and interpretable feature (i.e., matching signal between two texts, which is the essence of information retrieval). In this paper, we propose a novel method to improve the generalization of dense retrieval via capturing matching signal called BERM. Fully fine-grained expression and query-oriented saliency are two properties of the matching signal. Thus, in BERM, a single passage is segmented into multiple units and two unit-level requirements are proposed for representation as the constraint in training to obtain the effective matching signal. One is \textit{semantic unit balance} and the other is \textit{essential matching unit extractability}. Unit-level view and balanced semantics make representation express the text in a fine-grained manner. Essential matching unit extractability makes passage representation sensitive to the given query to extract the pure matching information from the passage containing complex context. Experiments on BEIR show that our method can be effectively combined with different dense retrieval training methods (vanilla, hard negatives mining and knowledge distillation) to improve its generalization ability without any additional inference overhead and target domain data.

\end{abstract}

\section{Introduction} \label{intro}

Dense retrieval encodes the texts to dense embeddings and efficiently gets the target texts via approximate nearest neighbor search~\cite{ann}. Compared with the traditional word-to-word exact matching methods such as BM25~\cite{bm25}, dense retrieval can capture the relevance at the semantic level of two texts. Because of the excellent performance in efficiency and effectiveness, dense retrieval has been widely used in first-stage retrieval that efficiently recalls candidate documents from the large corpus~\cite{dpr,ance}. 

\begin{figure}[t]
\centering
\includegraphics[width=\linewidth]{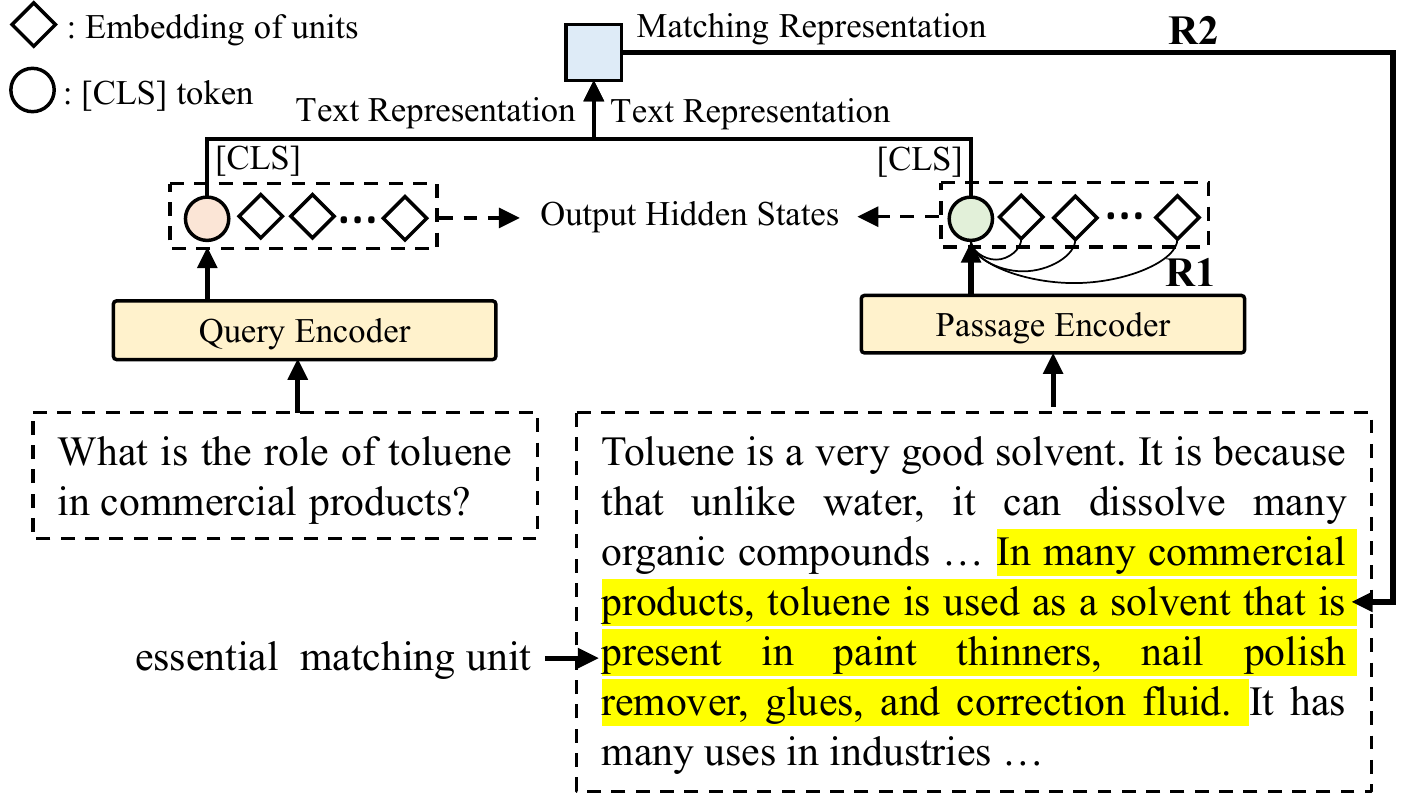} 
\caption{Idea of our method. \textbf{R1}: semantic unit balance. \textbf{R2}: essential matching unit extractability.}
\label{comparision}
\end{figure}

However, recent studies show that the excellent performance of dense retrieval relies on the training on large in-domain datasets. When the trained dense retrieval models are applied to the domains that are inconsistent with the training datasets (i.e., zero-shot setting), the performance of the models drops seriously~\cite{Examination,beir}. The poor generalization limits the application scenarios of dense retrieval because it is 
common that not enough training samples can be obtained in some domains such as medicine, biology and law that have restrictions on data privacy or require professional knowledge to annotate. 

In this work, we point out that according to out-of-domain generalization learning theory~\cite{ood}, making the model capture domain-invariant feature (i.e., essence of tasks) is effective in improving generalization ability. As for dense retrieval, matching signal between query and passage is the important domain-invariant feature and reflects the essence of information retrieval (IR). For example, MoDIR~\cite{MIDR} shows that representation from the interaction-based cross-encoder (more fine-grained description for matching) is much more domain-invariant than it from dense retrieval. Match-Prompt~\cite{matchprompt}, NIR-Prompt~\cite{nirprompt} and MatchPyramid~\cite{matchpyrimad} point out the positive significance of matching signals for various IR tasks. The challenge of making dense retrieval model learn to capture matching signal is that in many IR tasks such as open-domain question answering~\cite{open-domain-qa} and document retrieval~\cite{adhoc}, the content that matches the query is usually only a unit of the text. The description of matching signal needs to distinguish the matching and not matching information in the text and estimate the overall relevance. This requires the retrieval model to be able to evenly express each unit in the text and dynamically extract matching units through the interaction of the two text representations. However, the requirement on efficiency in first-stage retrieval makes dense retrieval only estimate relevance via vector similarity such as dot product and cosine. Previous training methods based on this architecture lack the above capability because of the coarse-grained training objective and interaction.

In this paper, we propose a novel method called BERM to capture the matching signal between query and passage, which is the domain-invariant feature, to improve the generalization ability of dense retrieval during the training on the single source domain without using the target domain data and other additional modules. First, we introduce a novel concept in dense retrieval, the matching representation. Matching representation is determined by the text representations (output of text encoder) of query and passage, which can reflect the matching information of query and passage. We propose that in the training of dense retrieval models, in addition to using contrastive loss~\cite{contrastive} to optimize the text representation, the information of the matching representation can be used as a constraint to assist the optimization. Based on this, we divide the single passage into multiple units (each sentence is a unit) and propose two requirements on the generalizable dense retrieval models as the constraint in training (shown in Figure~\ref{comparision}). One is \textit{semantic unit balance of text representation} (\textbf{R1}). The other is \textit{essential matching unit extractability of matching representation} (\textbf{R2}). These two requirements can be integrated into different dense retrieval training methods and address the challenge mentioned above. \textbf{R1} means the semantics of units in a passage are implicitly aggregated to its text representation and the text representation should evenly and comprehensively express the semantics of each unit. \textbf{R2} means that the combination of text representations of query and passage (i.e., matching representation) should extract the information of the matching (i.e, the text chunk in the passage that matches the query and we call it \textbf{essential matching unit}) while reducing the overfitting of domain biases. This reflects the ability of the dense retrieval model to determine and score the information that really matches the query in a passage containing complex context, which is the essence of the dense retrieval and domain-invariant. \textbf{R1} and \textbf{R2} achieve that on the premise that the text representation expresses each unit in a balanced manner, to make essential matching units for different queries be extracted, the semantics of units tend to be orthogonal to each other. In this way, in dot product between representations of query and passage, the semantics of essential matching unit are preserved, while the other units are masked, which is suitable for matching.

Experiments on the standard zero-shot retrieval benchmark (BEIR) show that our method can be effectively combined with different dense retrieval training methods (vanilla, hard negatives mining, and knowledge distillation) to improve the generalization ability without any additional modules, inference overhead, and target domain data. Even in domain adaptation, our method is also effective and performs better than baselines.

\section{Related Work}
Dense retrieval estimates the relevance via representations of two texts. DPR~\cite{dpr} combines dense retrieval with pre-trained models for open-domain question answering~\cite{open-domain-qa}. Besides, some methods focus on obtaining more valuable negatives~\cite{Rocketqa,ance,hard}. Some methods use a more powerful reranker for knowledge distillation~\cite{tas-b,dis1}. Recently, the generalization of dense retrieval has received attention. \cite{Examination} performs the examination of the generalization of dense retrieval. BEIR~\cite{beir} is proposed as the benchmark to evaluate the zero-shot ability of information retrieval models. MoDIR~\cite{MIDR} uses the data from source and target domains for adversarial training to perform unsupervised domain adaptation. GenQ~\cite{genq} and GPL~\cite{gpl} generate queries and pseudo labels for domain adaptation. Contriever~\cite{contriever} uses contrastive pre-training on large corpus (Wikipedia and CC-Net~\cite{ccnet}). COCO-DR~\cite{coco} performs unsupervised pre-training on target domain and introduces distributional robust optimization. GTR~\cite{gtr} scales up the model size to improve the generalization. ~\cite{hybrid,sparse} introduce sparse retrieval to achieve better generalization. Improvement of generalization of dense retrieval in previous studies comes from the adaptation of the target domain, knowledge from large pre-training corpus, and assistance of sparse retrieval but not dense retrieval itself. They need to obtain the target domain data in the training or increase the complexity of the system. In this paper, we introduce a novel method to improve the generalization of dense retrieval without target domain data and additional modules from the perspective of learning the generalizable representation for matching. Our method is general and can be combined with different dense retrieval training methods.

One thing must be emphasized that the methods of multi-view dense retrieval~\cite{mvr,dscr} also divide a passage into multiple units, but our method is essentially a completely different method. Multi-view dense retrieval uses multiple representations to fully express a passage from multiple views, which focuses on in-domain retrieval. Our method uses multiple units to make the model learn to extract essential matching unit from the passage containing complex context, which is domain-invariant for generalization. In our method, multiple units are only used as the constraint for optimization in training and only a single representation is used in inference. Learning-based sparse retrieval such as COIL~\cite{coil} and SPLADE~\cite{SPLADE} also aim to express fine-grained token-level semantics but they need multiple vectors to represent tokens in passage (COIL) or sparse-vector of vocabulary size (SPLADE) and calculates the score by token-to-token matching, which is not suitable for dense retrieval that uses single dense vector to perform representation and dot product.

\section{Motivation} \label{motivation}
Dense retrieval is hard to generalize to unseen domains due to its weak modeling of domain-invariant feature (i.e., matching signal between two texts, which is the essence of information retrieval). Fully fine-grained expression (\textbf{P1}) and query-oriented saliency (\textbf{P2}) are two properties of the matching signal. These two require the passage representation to be able to evenly express each unit in the text, and dynamically extract matching units according to the interaction with different queries. For example, BM25 uses one-hot to evenly express each word of the text, only scores matching words, and ignores not matching words through word-to-word exact matching of the two texts. Cross-encoder uses word embedding to represent the semantics of each token and uses attention to describe the token-to-token semantic matching between texts in a fine-grained manner. 

In this paper, based on the above two properties, for the training of dense retrieval, we segment a single passage into multiple units and propose two requirements as the constraint in training so that dense retrieval can capture the stronger matching signal and produces a suitable representation for matching. One is \textit{semantic unit balance of text representation} (\textbf{R1}), and the other is \textit{essential matching unit extractability of matching representation} (\textbf{R2}). 
\begin{figure}[t]
\centering
\includegraphics[width=\linewidth]{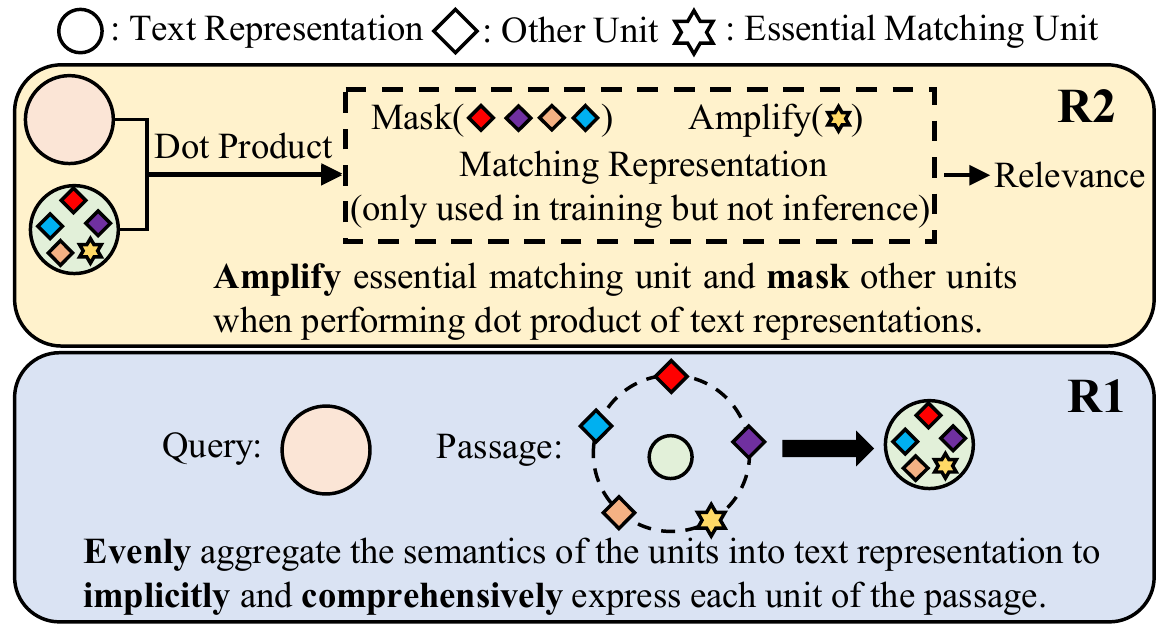} 
\caption{An illustration of the effects of \textbf{R1} and \textbf{R2}.}
\label{motivation_fig}
\end{figure}
Under \textbf{R1}, text representation evenly aggregates semantics of the units in the passage to comprehensively express the passage in a fine-grained manner. Besides, \textbf{R1} is the premise of \textbf{R2}. It is because that matching representation is composed of text representations from passage and query. Unbalanced semantic expression of different units in text representation will affect the identification of essential matching unit in matching representation because it leads to different preferences for different units. Under \textbf{R2}, essential matching unit for the query can be extracted from the passage and reflected in matching representation. Unlike using one-hot or word embedding to explicitly express the semantics of each unit and extract matching information through token-to-token interaction, as shown in Figure~\ref{motivation_fig}, \textbf{R1} makes the model implicitly aggregate the semantics of each unit into the text representation to satisfy \textbf{P1}, and \textbf{R2} makes the semantics of units tend to be orthogonal to each other (shown in Table~\ref{var-unit}). In dot product between representations of query and passage, semantics of essential matching unit are preserved, while the other units are masked, which can satisfy \textbf{P2}. Our method unlocks the ability of dense retrieval to capture matching signal without additional interaction.

\section{Our Method}
\begin{figure*}[t]
\centering
\includegraphics[width=\textwidth]{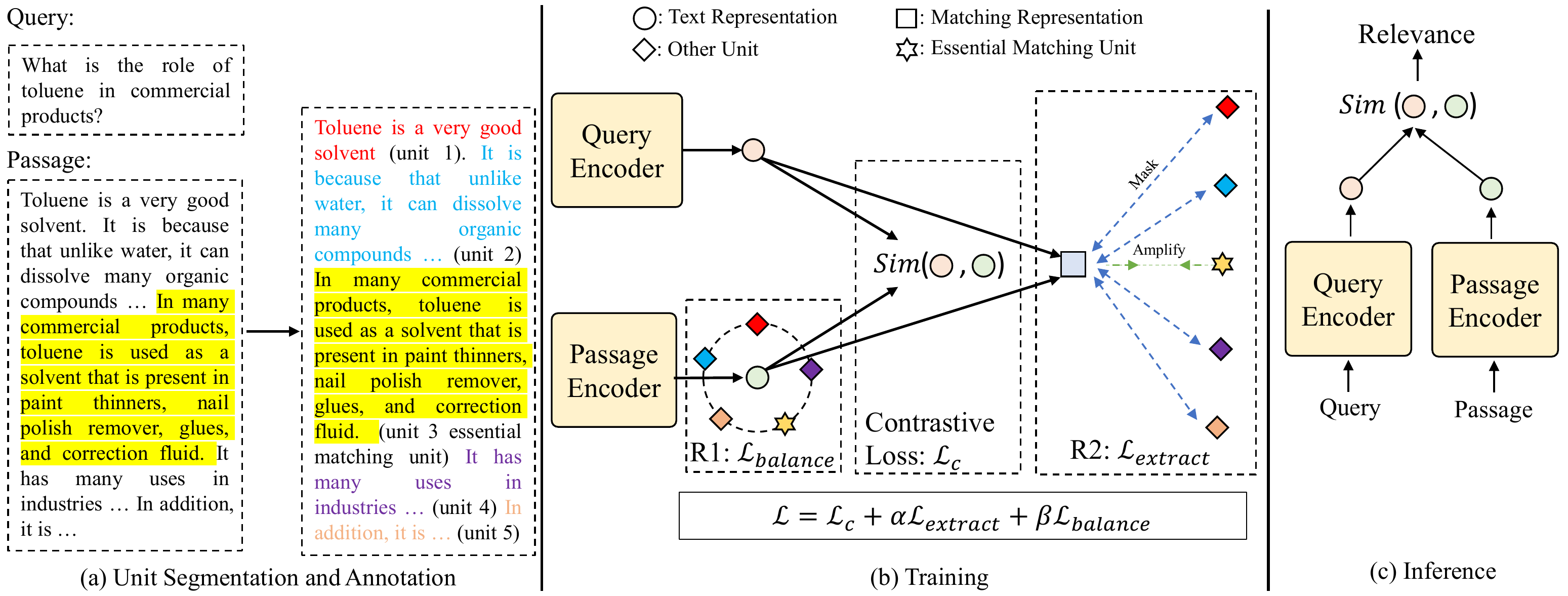} 
\caption{Overview of BERM. (a) Unit Segmentation and Annotation: Segment passage into multiple units (each sentence is a unit) and use the annotator to identify essential matching unit. (b) Training:  Training models under the constraints of two requirements on the generalizable dense retrieval for matching. (c) Inference: Same as the mainstream dense retrieval, without introducing additional inference and storage overhead.}
\label{overview}
\end{figure*}

This section introduces the implementation of our method (Figure~\ref{overview}). Our method optimizes the relationship between the representations and the units in the passage. Therefore, before training, we perform unit segmentation and annotate the essential matching unit for the datasets. Then, we design loss functions according to the requirements of \textbf{R1} and \textbf{R2} and combine these functions with task loss of dense retrieval (contrastive loss) for joint training.
\subsection{Unit Segmentation and Annotation} \label{unit seg}
Given each positive query-passage pair $(q,p_{pos})$ in training data, we segment positive passage into multiple units $U$ as shown in Figure~\ref{overview} (a) (We use the sentence as the segmentation granularity to ensure that each unit has complete semantic information.):
\begin{equation}
p_{pos} \stackrel{Segment}{\longrightarrow} U = \{u_1, u_2, ... , u_n\}.
\label{seg}
\end{equation}
For $U$ and $q$, BM25 is used to compute the word-to-word matching score $S_{bm25}$ between $q$ and $u_i \in U$:
\begin{equation}
    S_{bm25} = \{bm25(q,u_1), ... , bm25(q,u_n)\}. 
\label{e1} \nonumber
\end{equation}
For the datasets for question-answering, a trained reader model is additionally introduced to compute the semantic matching score $S_{reader}$ between $q$ and $u_i \in U$. Specifically, reader model computes the probability distribution $A = \{a_1, a_2, .., a_t\}$ of the starting positions of the answer in $p_{pos}$. $a_i$ indicates the probability that the $i\mbox{-}th$ token in $p_{pos}$ is the starting of the answer to $q$. For each $u_i \in U$, the semantic matching score from the reader model is:
\begin{equation}
r_i = max(A{[s_{u_i} : d_{u_i}]}),
\label{tokens} 
\end{equation}
where $[s_{u_i} : d_{u_i}]$ are the indexes of tokens in $u_i$. The hybrid matching score $h_i$ between $u_i$ and $q$ is:
\begin{equation}
h_i = bm25(q,u_i) + \delta r_i,
\label{e1} \nonumber
\end{equation}
where $\delta$ is a hyperparameter. We set $\delta$ as $0.1$ to give BM25 a higher weight than the reader. It is because the word-to-word exact matching of BM25 is more domain-invariant and conducive to generalization than the semantic matching of reader~\cite{beir}. Then we get matching score list $H=\{h_1, h_2, ... , h_n\}$ for $U$. The essential matching unit is the unit corresponding to the maximum value in $H$. For the pair $(q,p_{pos})$, $y_i$ in label list $Y=\{y_1, ... , y_n\}$ for essential matching unit is that if $i$ is the index corresponding to the maximum value in $H$, $y_i=1$, otherwise, $y_i=0$.

\subsection{Training for Generalization} \label{training}
Based on the analysis of properties of matching signal in Section~\ref{motivation}, we propose two requirements as the constraints in the training of dense retrieval to get a generalizable representation for matching (shown in Figure~\ref{overview} (b)). These two requirements enable dense retrieval to extract essential matching information under the premise of balanced expression of each unit, so as to learn domain-invariant feature (i.e., matching signal) for generalization. 

\textbf{Implementation of R1.} 
The first requirement is \textit{semantic unit balance of text representation}, which means that the text representation of the passage encoder can comprehensively express the semantics of each unit in a balanced manner. Given the passage $p_{pos}$, the text encoder $g(\cdot;\theta)$, output hidden states $\boldsymbol{Z}=g(p_{pos};\theta)$. Text representation $\boldsymbol{t_{p}}$ of $p_{pos}$ is the embedding of [CLS] token of $\boldsymbol{Z}$. The embeddings $\boldsymbol{E}$ of units in $p_{pos}$ can be obtained from $\boldsymbol{Z}$ as the segmentation in Equ.(\ref{seg}):
\begin{equation}
\boldsymbol{E} = \{\boldsymbol{e_1}, \boldsymbol{e_2}, ..., \boldsymbol{e_n}\},
\label{emb_u} 
\end{equation}
where $\boldsymbol{e_i}$ is the embedding of the corresponding unit $(u_i)$ and it is the average pooling of the embeddings of tokens ($\boldsymbol{Z}[s_{u_i}:d_{u_i}]$) in the unit, where $[s_{u_i} : d_{u_i}]$ are the indexes of tokens in $u_i$. Under the constraint of \textbf{R1}, the relationship between $\boldsymbol{t_p}$ and $\boldsymbol{E}$ is described by the loss function as:
\begin{equation}
\mathcal{L}_{balance} = D_{KL}[\boldsymbol{b}||sim(\boldsymbol{t_p},\boldsymbol{E})],
\label{balance} 
\end{equation}
where $D_{KL}[\cdot||\cdot]$ is KL-divergence loss, $\boldsymbol{b}=[\frac{1}{n},...,\frac{1}{n}]$ is a uniform distribution with equal values and $sim(\boldsymbol{t_p},\boldsymbol{E})=\{dot(\boldsymbol{t_p},\boldsymbol{e_i})|\boldsymbol{e_i}\in \boldsymbol{E}\}$ is a distribution to represent the semantic similarity between $\boldsymbol{t_p}$ and $\boldsymbol{e_i} \in \boldsymbol{E}$, $dot(\cdot,\cdot)$ is dot product.

\textbf{Implementation of R2.} \label{r2}
The second requirement is \textit{essential matching unit extractability of matching representation}, which means that under the premise of \textbf{R1}, matching representation can saliently represent the unit where the essential matching block is located in. The motivation for this design is discussed in Section~\ref{intro} and~\ref{motivation}. Given the positive query-passage pair $(q,p_{pos})$, text encoder $g(\cdot;\theta)$, and the text representations for $q$ ($\boldsymbol{t_q}$) and $p_{pos}$ ($\boldsymbol{t_p}$). Matching representation $\boldsymbol{m} \in \mathbb{R}^{v}$ ($v$ is the dimension of representation) for $q$ and $p_{pos}$ can be obtained by the combination of $\boldsymbol{t_q} \in \mathbb{R}^{v}$ and $\boldsymbol{t_p} \in \mathbb{R}^{v}$ as:
\begin{equation}
\boldsymbol{m} = \textit{GELU}(\boldsymbol{t_q} \odot \boldsymbol{t_p}),
\label{e1} \nonumber
\end{equation}
where $\odot$ is the element-wise multiplication operator, and $\textit{GELU}(\cdot)$ is activation function~\cite{gelu} to introduce random regular. Under the premise of \textbf{R1}, $\boldsymbol{t_p}$ can express the semantics of the units in $p_{pos}$ in a balanced manner. In addition, the semantic representation of essential matching unit is more similar to $\boldsymbol{t_q}$ than other units because it really matches the query $q$. Based on this, the model can be trained to achieve the goal that element-wise multiplication between $\boldsymbol{t_q}$ and $\boldsymbol{t_p}$ can amplify similar patterns (i.e, semantic representation of essential matching unit) and mask the signals of other context units. This design can be supported by convolutional neural networks~\cite{cnn1} whose convolution operation can amplify similar patterns in tensors~\cite{cnn2}. For the $p_{pos}$, different $q$ amplifies different matching units, which makes $\boldsymbol{m}$ reflect the semantics of the corresponding essential matching unit. Besides, $\boldsymbol{m}$ is obtained by element-wise multiplication between $\boldsymbol{t_q}$ and $\boldsymbol{t_p}$, which is an important part of estimating the relevance of two texts because $ dot(\boldsymbol{t_q},\boldsymbol{t_p}) = sum(\boldsymbol{t_q} \odot \boldsymbol{t_p})$. Thus, the optimization of $\boldsymbol{m}$ can enable the model to obtain the ability to extract essential matching unit according to different queries when estimating relevance. In training, our method utilizes the cross-entropy loss function to optimize the semantic distance between $\boldsymbol{m}$ and each unit to identify the corresponding essential matching units. Given query-passage pair ($q$, $p_{pos}$), the embeddings $\boldsymbol{E}$ of the units in $p_{pos}$ as described in Equ. (\ref{emb_u}), and the label $Y$ for essential matching unit of ($q$, $p_{pos}$) as described in Sec.~\ref{unit seg}. Loss function for \textbf{R2} is:
\begin{equation}
\mathcal{L}_{extract} = -\sum_{i=1}^{n}y_i \log(dot(\boldsymbol{m},\boldsymbol{e_i})),
\label{extract} 
\end{equation}
where $\boldsymbol{e_i} \in \boldsymbol{E}$, $y_i \in Y$. $\boldsymbol{m}$ is only used as the constraint in training but has important implications for inference. It is because that $\boldsymbol{m}$ is the combination of text representations ($\boldsymbol{t_p}$ and $\boldsymbol{t_q}$). The optimization for $\boldsymbol{m}$ is training the text encoder to output the text representation that is suitable for matching to improve the generalization ability.

\textbf{Effect of \textbf{R1} and \textbf{R2}.} Table~\ref{var-unit} indicates that compared with previous dense retrieval methods, our method makes the semantics of units in text representation tend to be orthogonal to each other. In dot product between two texts, semantics of essential matching unit are preserved, while the other units are masked to capture matching signal.

\textbf{Total Loss.}
In addition to $\mathcal{L}_{extract}$ and $\mathcal{L}_{balance}$, contrastive loss is used to train the dense retrieval model~\cite{dpr} as:
\begin{equation}
\mathcal{L}_{c} = -{\exp(dot(\boldsymbol{t_q},\boldsymbol{t_{p^+}})) \over \exp(dot(\boldsymbol{t_q},\boldsymbol{t_{p^+}}))+\exp(dot(\boldsymbol{t_q},\boldsymbol{t_{p^-}}))}.
\label{con} \nonumber
\end{equation}
So the total loss for training in our method is:
\begin{equation}
\mathcal{L} = \mathcal{L}_{c} + \alpha \mathcal{L}_{extract} + \beta \mathcal{L}_{balance},
\label{con} \nonumber
\end{equation}
where $\alpha$ and $\beta$ are the hyperparameters.

\section{Experiments}
This section introduces the experimental setups and analyzes the results.
\subsection{Experimental Setups}
\quad \textbf{Datasets.} We use MS-MARCO~\cite{msmarco} as the training data (source domain) and choose the 14 publicly available datasets from BEIR~\footnote{The left four are unavailable due to copyright restrictions.}, a heterogeneous benchmark to evaluate the generalization ability of retrieval models. In addition, we also introduce OAG-QA~\cite{oag-qa} to evaluate the topic generalization ability. Details of datasets are in Appendix~\ref{Datasets}.

\begin{table*}[]
\setlength\tabcolsep{8pt}
\centering
\scalebox{0.85}{
\begin{tabular}{l|c|cc|cc|cc}
\hline
\multirow{2}{*}{Datasets} & Jaccard Sim & \multicolumn{2}{c|}{Vanilla}    & \multicolumn{2}{c|}{Knowledge Distillation} & \multicolumn{2}{c}{Hard Negatives} \\
                          & Unigrams    & DPR            & DPR+BERM        & KD                   & KD+BERM               & ANCE             & ANCE+BERM        \\ \hline
SciFact                   & 22.16       & 0.478          & \ \textbf{0.495$^{\dag}$} & 0.481                & \ \textbf{0.504$^{\dag}$}       & 0.507            & \ \textbf{0.511$^{\dag}$}  \\
NFCorpus                  & 23.45       & 0.208          & \ \textbf{0.234$^{\dag}$} & 0.205                & \ \textbf{0.242$^{\dag}$}       & 0.237            & \ \textbf{0.248$^{\dag}$}  \\
TREC-COVID                & 26.80       & 0.561          & \ \textbf{0.600$^{\dag}$} & 0.490                & \ \textbf{0.505$^{\dag}$}       & 0.654            & \ \textbf{0.661$^{\dag}$}  \\
SCIDOCS                   & 27.92       & 0.108          & \ \textbf{0.120$^{\dag}$} & 0.111                & \ \textbf{0.115$^{\dag}$}       & 0.122            & \ \textbf{0.130$^{\dag}$}  \\
DBPedia                   & 30.16       & 0.236          & \ \textbf{0.256$^{\dag}$} & 0.245                & \ \textbf{0.264$^{\dag}$}       & 0.281            & \ \textbf{0.293$^{\dag}$}  \\
CQADupStack               & 30.64       & \textbf{0.281} & 0.279          & \textbf{0.290}       & 0.281                & \textbf{0.296}   & 0.290           \\
HotpotQA                  & 30.87       & 0.371          & \ \textbf{0.386$^{\dag}$} & 0.427                & \ \textbf{0.438$^{\dag}$}       & 0.456            & \ \textbf{0.463$^{\dag}$}  \\
ArguAna                   & 32.92       & 0.414          & \ \textbf{0.435$^{\dag}$} & 0.435                & \ \textbf{0.437$^{\dag}$}       & 0.415            & \ \textbf{0.428$^{\dag}$}  \\
Climate-FEVER             & 34.79       & 0.176          & \ \textbf{0.187$^{\dag}$} & 0.189                & \ \textbf{0.195$^{\dag}$}       & 0.198            & \ \textbf{0.201$^{\dag}$}  \\
FEVER                     & 34.79       & \textbf{0.589} & 0.585          & 0.633                & \ \textbf{0.664$^{\dag}$}       & 0.669            & \ \textbf{0.674$^{\dag}$}  \\
FiQA-2018                 & 35.95       & \textbf{0.275} & 0.272          & \textbf{0.286}       & 0.285                & \textbf{0.295}   & 0.287           \\
Tóuche-2020               & 37.02       & 0.208          & \textbf{0.210$^{\dag}$}          & 0.215       & \textbf{0.216$^{\dag}$}                & 0.240            & \ \textbf{0.248$^{\dag}$}  \\
Quora                     & 39.75       & 0.842          & \ \textbf{0.853$^{\dag}$} & 0.832                & \ \textbf{0.836$^{\dag}$}       & 0.852            & \ \textbf{0.854$^{\dag}$}  \\
NQ                        & 47.27       & \textbf{0.398} & 0.394          & \textbf{0.420}       & 0.419                & 0.446            & \ \textbf{0.450$^{\dag}$}  \\ \hdashline
Avg                      & -           & 0.368          & \ \textbf{0.379} & 0.376                & \textbf{0.386}       & 0.405            & \textbf{0.410}  \\ \hline
\end{tabular}
}
  \caption{Zero-shot performance on BEIR (nDCG@10) without any target domain data. \textbf{Bold} indicates the better performance in the same training method. ${\dag}$: results with significant performance improvement with p-value $ \leq 0.05$ compared with baselines. Datasets are ordered by the Jaccard similarity between the source domain (MS-MARCO).}
\label{main}
\end{table*}

\textbf{Baselines.}
Our method (BERM) aims to improve the generalization of dense retrieval without any additional modules and target domain data, and it can be combined with different dense retrieval training methods. We select three mainstream dense retrieval training methods including vanilla, hard negatives mining, and knowledge distillation as the baselines. We follow DPR~\cite{dpr} to perform vanilla, follow ANCE~\cite{ance} to perform hard negatives mining and use a trained cross-encoder as the teacher model to perform knowledge distillation. We compare the change in generalization after combining BERM with these three methods to show the effectiveness of our method. Besides, as previous methods need to obtain target domain data for domain adaptation such as MoDIR~\cite{MIDR}, GenQ~\cite{genq}, GPL~\cite{gpl} and COCO-DR~\cite{coco}, we also compare our method with these methods in domain adaptation setting. Details of baselines are in Appendix~\ref{Baseline}.

\textbf{Implementation Details.}
To maintain a fair comparison, we follow~\cite{ance} to keep all common hyperparameters (learning rate and batch size, etc.) the same as the three dense retrieval training methods in the baselines. The model is initialized by Roberta$_{base}$ 125M. For the hyperarameters in BERM, $\delta$ is 0.1, $\alpha$ is 0.1 and $\beta$ is 1.0. In domain adaptation, we combine BERM with continuous contrastive pretraining~\cite{coco} to perform unsupervised pre-training on BEIR and use BERM to fine-tune the model on MS-MARCO. We train the model with Pytorch~\cite{pytorch} and Hugging Face~\cite{huggingface} on 2 Tesla V100 32GB GPUs for about 72 hours.

\subsection{Retrieval Performance}
\quad \textbf{Main Results.} Table~\ref{main} shows the main results on BEIR of different dense retrieval training methods. The results indicate that our method (BERM) can be combined with three mainstream dense retrieval training methods (vanilla, knowledge distillation, and hard negatives) to improve the generalization ability without any additional modules and target domain data. For a fair comparison, we combine BERM with the baselines and ensure that their common hyperparameters are consistent. We compute the Jaccard similarity~\cite{jaccard} between each dataset and MS-MARCO, which can reflect the domain shift between the source and target domain. Table~\ref{main} shows that our method is more effective for the datasets with lower Jaccard similarity between MS-MARCO (i.e., domain shift is more significant). This result reflects the ability of our method to capture domain-invariant feature. DPR+BERM and KD+BERM are better than KD, which shows that BERM more effectively enables dense retrieval to learn to capture matching signal than knowledge distillation from cross-encoder.

\textbf{Topic Generalization.} Table~\ref{oag} shows the generalization performance of DPR and DPR+BERM on different topics of QAG-QA. Topic generalization is important for out-of-domain generalization, which reflects the availability of dense retrieval model for topics with different word distributions. The results show that BERM can significantly improve cross-topic generalization of dense retrieval. 
\begin{table}[t]
\setlength\tabcolsep{8pt}
\centering
\scalebox{0.85}{
\begin{tabular}{lcc}
\toprule
Topic                    & \multicolumn{1}{c}{DPR} & \multicolumn{1}{c}{DPR+BERM} \\ \hline
Geometry                 & 0.324                   & \textbf{0.343$^{\dag}$}              \\
Mathematical statistics  & 0.238                   & \textbf{0.246$^{\dag}$}              \\
Polynomial               & 0.174                   & \textbf{0.209$^{\dag}$}              \\
Calculus                 & 0.198                   & \textbf{0.207$^{\dag}$}              \\
Number theory            & 0.268                   & \textbf{0.281$^{\dag}$}              \\
Matrix                   & 0.259                   & \textbf{0.296$^{\dag}$}              \\
Black hole               & 0.107                   & \textbf{0.143$^{\dag}$}              \\
Classical mechanics      & 0.209                   & \textbf{0.242$^{\dag}$}              \\
Physical chemistry       & 0.154                   & \textbf{0.183$^{\dag}$}              \\
Biochemistry             & 0.306                   & \textbf{0.333$^{\dag}$}              \\
Health care              & 0.389                   & \textbf{0.401$^{\dag}$}              \\
Evolutionary biology     & 0.294                   & \textbf{0.316$^{\dag}$}              \\
Cognitive neuroscience   & 0.303                   & \textbf{0.310$^{\dag}$}              \\
Algorithm                & 0.266                   & \textbf{0.271$^{\dag}$}              \\
Neural network           & 0.179                   & \textbf{0.191$^{\dag}$}              \\
Data mining              & 0.291                   & \textbf{0.336$^{\dag}$}              \\
Computer graphics images & 0.255                   & \textbf{0.277$^{\dag}$}              \\
Optimization             & 0.230                   & \textbf{0.244$^{\dag}$}              \\
Linear regression        & 0.153                   & \textbf{0.189$^{\dag}$}              \\
Economics                & 0.299                   & \textbf{0.332$^{\dag}$}    \\ \toprule        
\end{tabular}
}
  \caption{Zero-shot performance on OAG-QA (metric is Top-20). \textbf{Bold}: better. ${\dag}$: significant improvement.}
\label{oag}
\end{table}

\textbf{Domain Adaptation.} Table~\ref{domain adapatation} shows that BERM achieves the best performance in domain adaptation compared with previous baselines. Specifically, BERM achieves the best average out-of-domain adaptation and in-domain performance. Besides, it gets the best dense retrieval results on seven datasets of BEIR, which is the most of all methods. Our method not only learns the word distribution of the target domain, but also learns the representation suitable for matching for the documents in the target corpus during domain adaptation.

\begin{table*}[t]
\setlength\tabcolsep{6pt}
\centering
\scalebox{0.85}{
\begin{tabular}{l|c|c|cccccc}
\hline
\multirow{2}{*}{Datasets} & Sparse                     & Late-Inter.    & \multicolumn{5}{c}{Dense}                                                        \\
                          & BM25                       & ColBERT        & MoDIR       & Contriever     & GenQ   & GPL        & COCO-DR        & BERM (ours) \\ \hline
MS-MARCO                  & 0.228                      & 0.401          & 0.388       & 0.407          & 0.408     & -     & 0.419          & 0.421  \\ \hline
SciFact                   & 0.665                      & 0.671          & 0.502       & 0.677          & 0.644     & 0.674     & {0.709}    & \textbf{ 0.720$^{\dag}$}  \\
NFCorpus                  & 0.325                      & 0.305          & 0.244       & 0.328          & 0.319     & 0.345     & {0.355}    & \textbf{ 0.357$^{\dag}$}  \\
TREC-COVID                & 0.656                      & 0.677          & 0.676       & 0.596          & 0.619     & 0.700     & {0.789}    & \textbf{ 0.795$^{\dag}$}  \\
SCIDOCS                   & 0.158                      & 0.145          & 0.124       & 0.165 & 0.143     & \textbf{0.169}     & 0.160       & 0.161     \\
DBPedia                   & 0.313                      & 0.392 & 0.284       & \textbf{0.413}          & 0.328     & 0.384     & {0.391}    & 0.391           \\
CQADupStack               & 0.299                      & 0.350          & 0.297       & 0.345          & 0.347     & 0.357     & {0.370}    & \textbf{ 0.374$^{\dag}$}  \\
HotpotQA                  & 0.603                      & 0.593          & 0.462       & \textbf{0.638} & 0.534     & 0.582     & {0.616}    & 0.610           \\
ArguAna                   & 0.414                      & 0.233          & 0.418       & 0.446          & 0.493     & \textbf{0.557}    & 0.493 & {0.490}     \\
Climate-FEVER             & 0.213                      & 0.184          & 0.206       & \textbf{0.237} & 0.175     & 0.235     & 0.211          & { 0.220$^{\dag}$}     \\
FEVER                     & 0.753                      & 0.771 & 0.680       & 0.758          & 0.669     & 0.759     & 0.751          & \textbf{ 0.760$^{\dag}$}     \\
FiQA-2018                 & 0.236                      & { 0.317}    & 0.296       & 0.329 & 0.308        & \textbf{0.344}  & 0.307          & 0.301           \\
Tóuche-2020               & 0.367             & 0.202          & \textbf{ 0.315} & 0.230          & 0.182        & 0.255  & 0.238          & 0.235           \\
Quora                     & 0.789                      & 0.854          & 0.856       & 0.865          & 0.830     & 0.836     & {0.867}    & \textbf{ 0.870$^{\dag}$}  \\
NQ                        & 0.329                      & 0.524 & 0.442       & 0.498          & 0.358     & 0.483     & 0.505          & \textbf{0.506}     \\ \hdashline
Avg w/o MS-MARCO                      & \multicolumn{1}{l|}{0.437} & 0.444          & 0.414       & 0.466          & 0.425     & 0.477     & 0.483          & \textbf{0.485}  \\ \hline
\end{tabular}
}
\caption{nDGC@10 performance on BEIR. \textbf{Bold} indicates the best dense retrieval performance.  ${\dag}$: results with significant performance improvement with p-value $ \leq 0.05$ compared with COCO-DR. Dense retrieval models are trained in the domain adaptation setting that unsupervised target domain data can be obtained.}
\label{domain adapatation}
\end{table*}

\subsection{Ablation Study} 
\quad \textbf{Influence of Loss Functions.} Table~\ref{ablation} shows the ablation study on the loss functions constrained by \textbf{R1} and \textbf{R2} via average performance on BEIR. The results indicate that without $\mathcal{L}_{balance}$, $\mathcal{L}_{extract}$ can not improve the generalization, which supports our intuition in Section~\ref{motivation} that only based on the balanced semantic expression of each unit in the text representation, the matching representation is meaningful for extracting the essential semantic unit. This experiment shows that the generalization can be improved significantly when the model is constrained by both \textbf{R1} and \textbf{R2}.

\begin{table}[H]
\centering
\setlength\tabcolsep{3pt}
\scalebox{0.8}{
\begin{tabular}{lccc}
\toprule
         & DPR+BERM & KD+BERM & ANCE+BERM \\ \hline
         & 0.379   & 0.386  & 0.410    \\ \hline
w/o $\mathcal{L}_{balance}$  & 0.365   & 0.371  & 0.392    \\
w/o $\mathcal{L}_{extract}$ & 0.372   & 0.383  & 0.406    \\ \toprule
\end{tabular}}
\caption{Ablation study on $\mathcal{L}_{balance}$ and $\mathcal{L}_{extract}$.}
\label{ablation}
\end{table}

\begin{figure}[H]
    \centering
    \subfigure[Effect of $\alpha$.]{
	\includegraphics[width=1.42in]{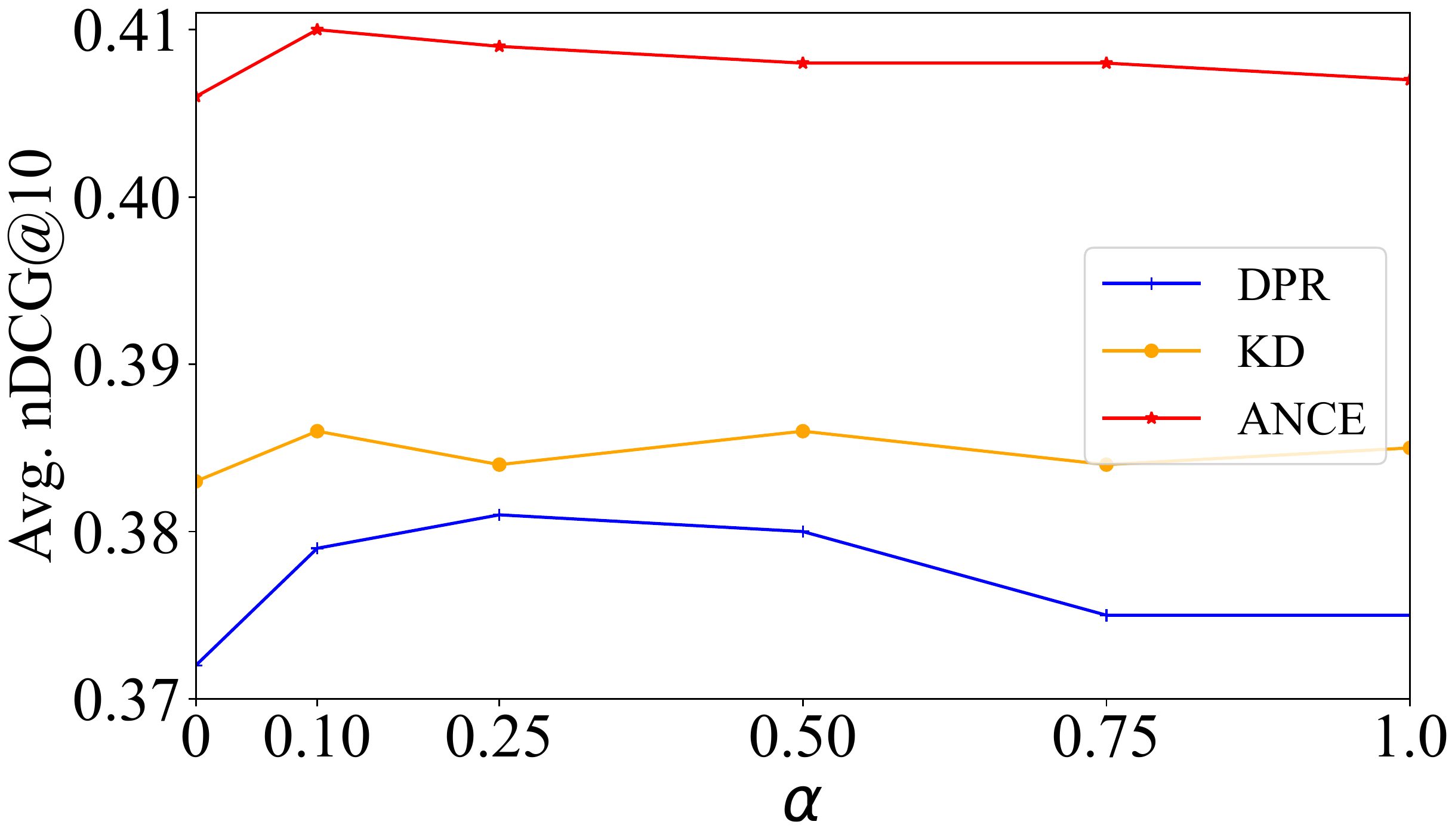}
	\label{tr_dpr}
    }
    \subfigure[Effect of $\beta$.]{
        \includegraphics[width=1.42in]{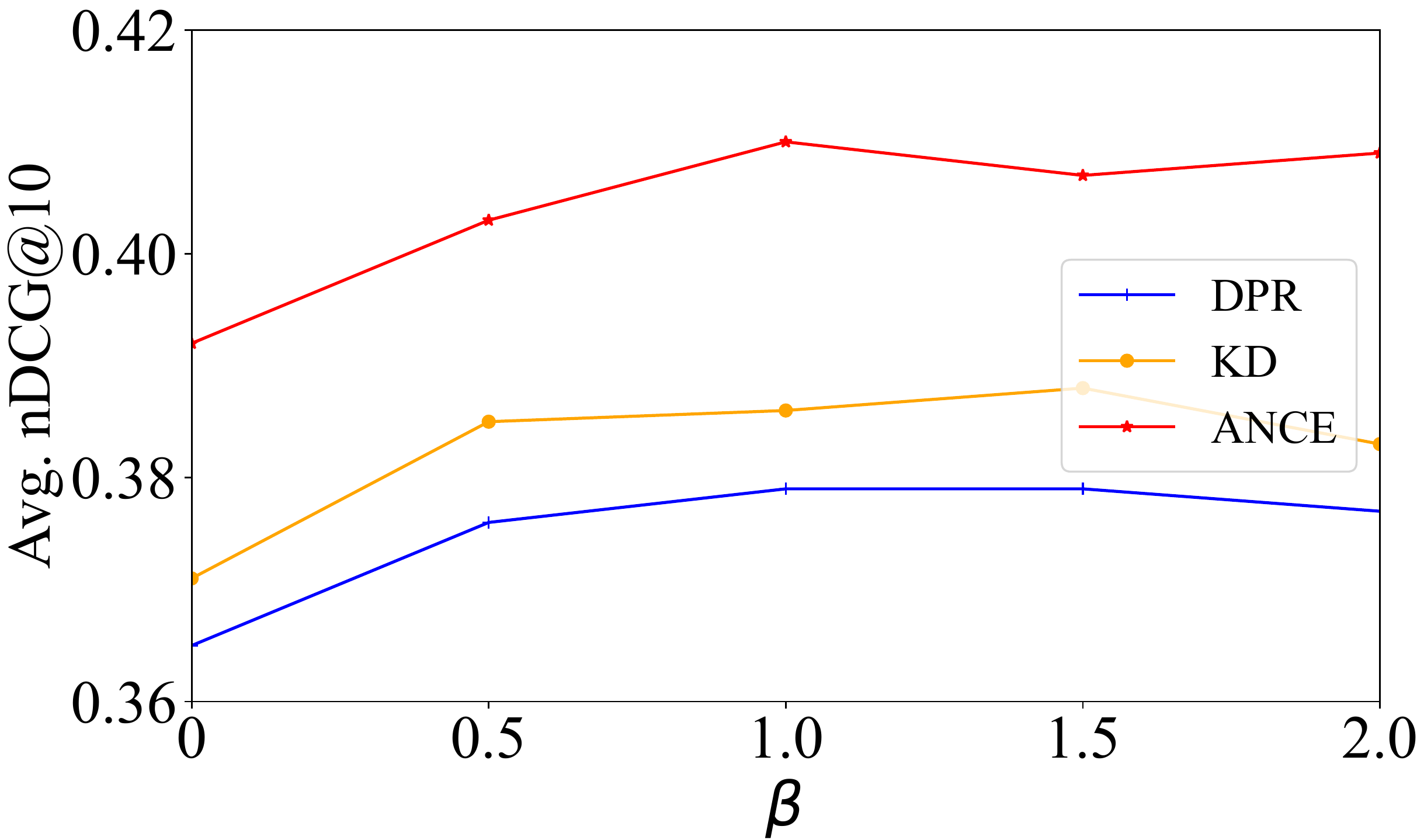}
    \label{tr_BERM}
    }
\caption{Performance varies with $\alpha$ and $\beta$.}
\label{hyparamters}
\end{figure}

\textbf{Influence of Hyperparameters.} Figure~\ref{hyparamters} shows the average nDCG@10 performance on BEIR with different $\alpha$ and $\beta$ that are used to tune the weights of different loss functions in training. When $\alpha$ is $0.1$ and $\beta$ is $1.0$, our method can achieve the best performance. When $\alpha$ and $\beta$ are too big, they will interfere with the optimization of the contrastive loss leading to performance degradation.

\subsection{Model Analysis} \label{analysis}
\begin{figure*}[t]
    \centering
    \subfigure[Text Rep.]{
	\includegraphics[width=1.47in]{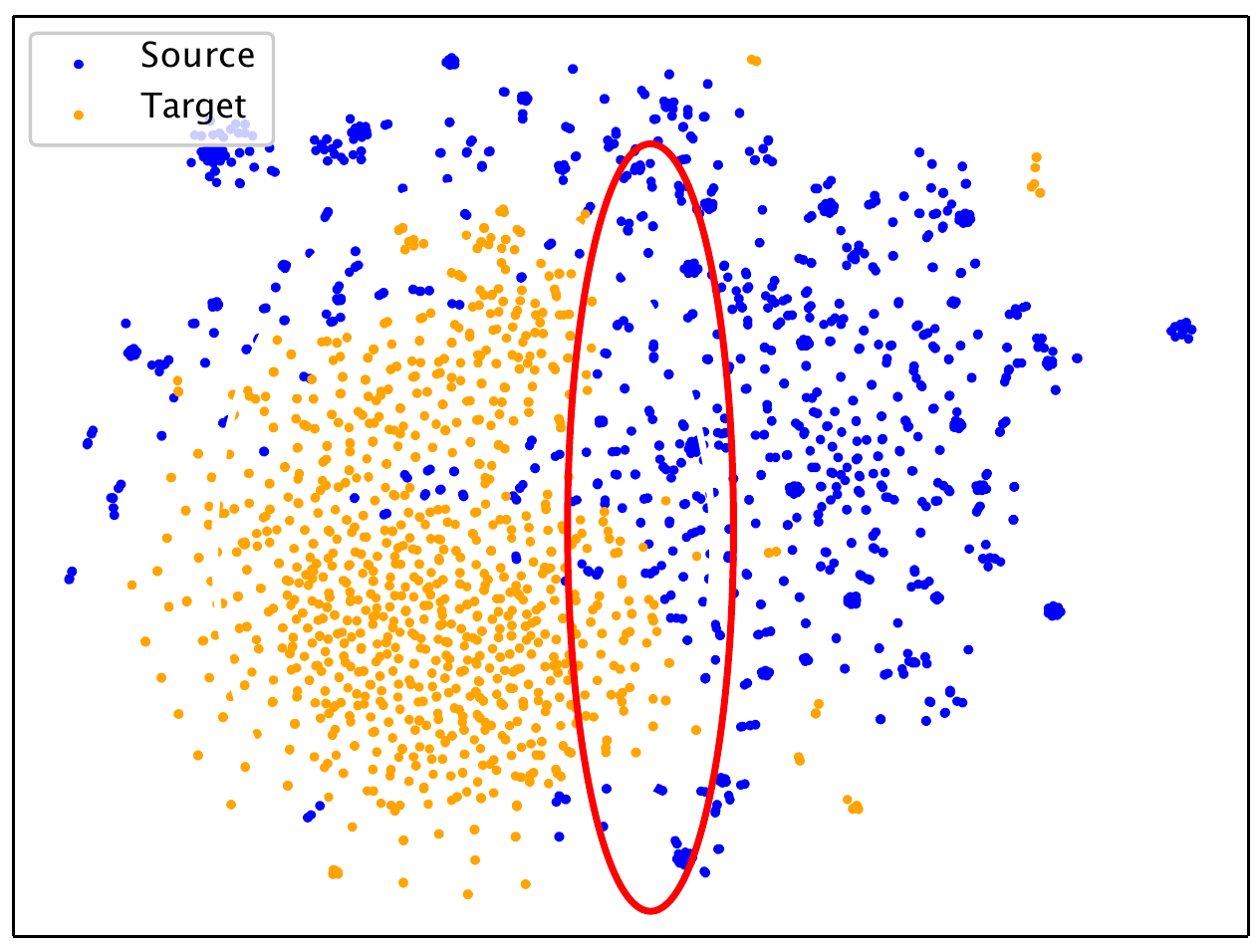}
	\label{tr_dpr}
    }
    \subfigure[Text Rep. (ours)]{
        \includegraphics[width=1.47in]{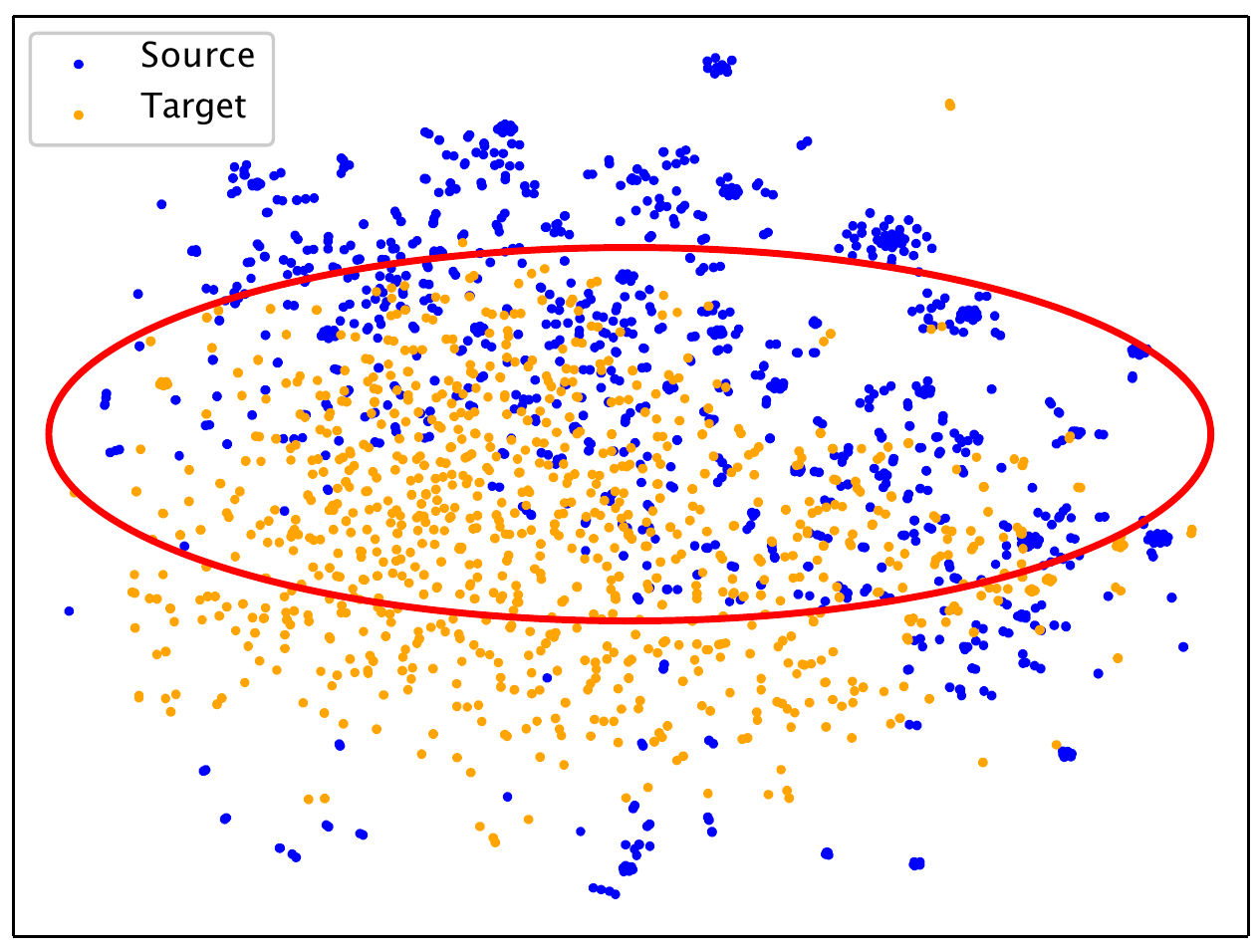}
    \label{tr_BERM}
    }
    \subfigure[Matching Rep.]{
        \includegraphics[width=1.47in]{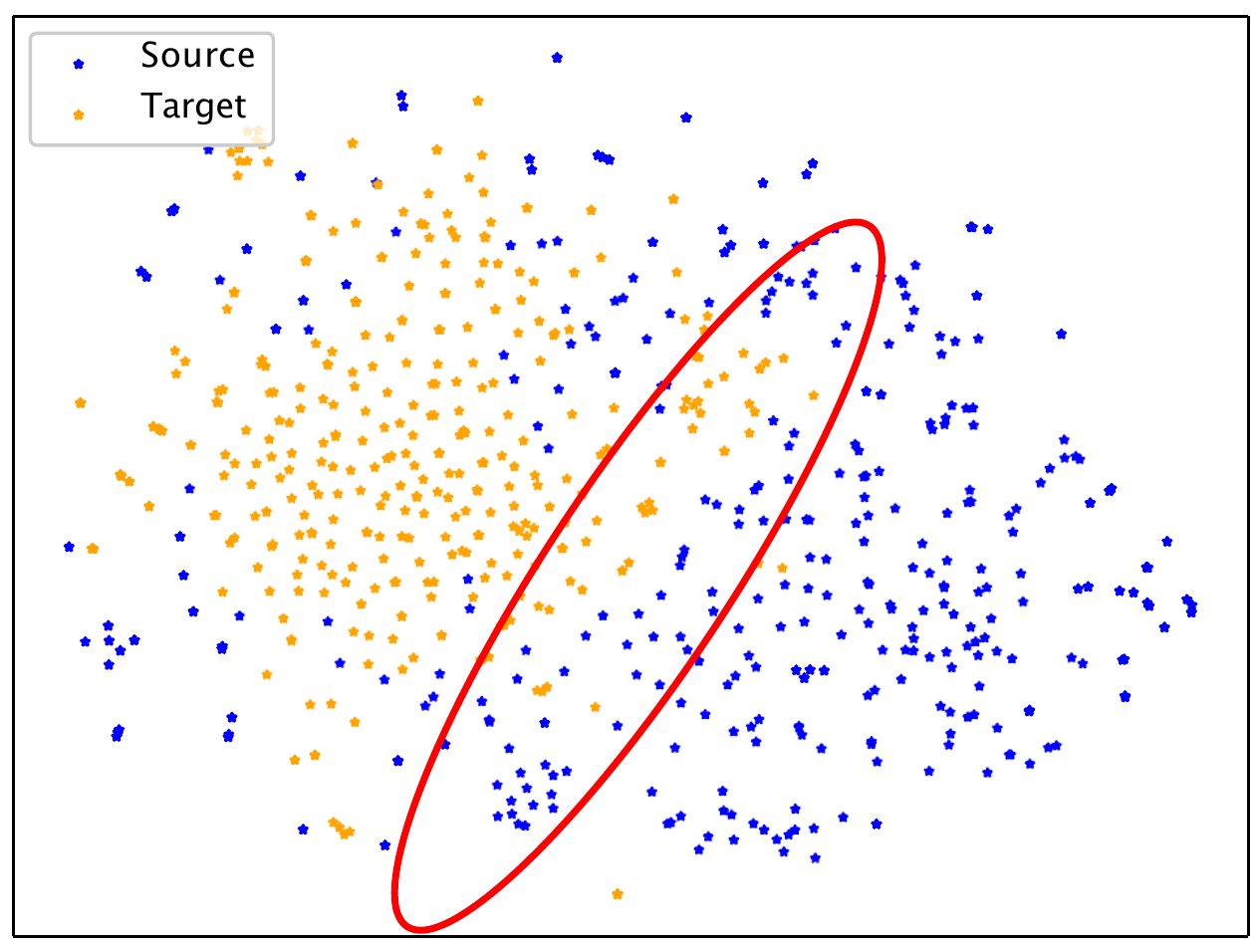}
    \label{mr_dpr}
    }
    \subfigure[Matching Rep. (ours)]{
        \includegraphics[width=1.47in]{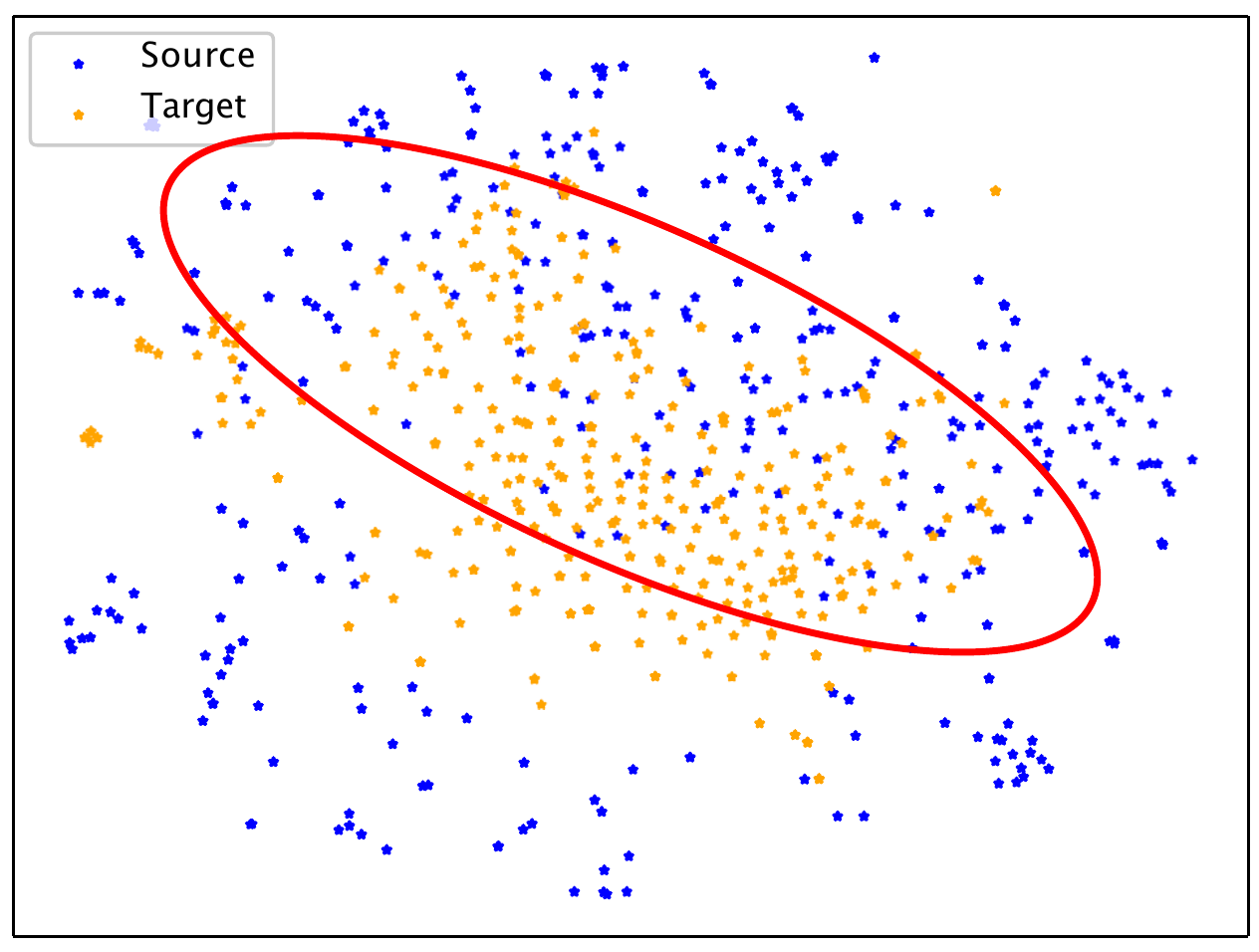}
    \label{mr_BERM}
    }
\caption{T-SNE of the text and matching representations for source and target domains.}
\label{domain-invariant}
\end{figure*}

\quad \textbf{Domain-invariant Representation.} Figure~\ref{domain-invariant} shows that our method is effective in capturing the domain-invariant feature of the representation. We utilize T-SNE to visualize the representations of source and target (SciFact) domains encoded by DPR and DPR+BERM respectively. The results indicate that representations of the two domains encoded by DPR are more separable. After combining our method, the two domains become more difficult to separate, which indicates that our method is more invariant to represent the texts in different domains. More datasets are in Appendix~\ref{domain-figure}.

\begin{table}[t]
\centering
\setlength\tabcolsep{4.5pt}
\scalebox{0.85}{
\begin{tabular}{lllllll}
\toprule
         & \multicolumn{2}{c}{DPR}                             & \multicolumn{2}{c}{KD}           & \multicolumn{2}{c}{ANCE}                            \\
Metric          & \multicolumn{1}{c}{Var.} & \multicolumn{1}{c}{Acc.} & \multicolumn{1}{c}{Var.} & Acc.  & \multicolumn{1}{c}{Var.} & \multicolumn{1}{c}{Acc.} \\ \hline
Baseline & 3.756                    & 0.407                    & 3.891                    & 0.415 & 3.432                    & 0.450                    \\
BERM      & 0.005                    & 0.778                    & 0.007                    & 0.803 & 0.003                    & 0.846       \\ \toprule            
\end{tabular}}
\caption{Variance of the semantic similarity between text representation and units (smaller the better). Accuracy to identify essential matching unit (bigger the better).}
\label{var}
\end{table}

\textbf{Evaluation of \textbf{R1} and \textbf{R2}.} Table~\ref{var} shows the effectiveness of \textbf{R1} and \textbf{R2}. We randomly sample 100,000 query-passage pairs from the test set. For each passage $p$, we compute semantic similarity between text representation and each unit via $sim(\boldsymbol{t_p},\boldsymbol{E})=\{dot(\boldsymbol{t_p},\boldsymbol{e_i})|\boldsymbol{e_i}\in \boldsymbol{E}\}$. We compute the variance of $sim(\boldsymbol{t_p},\boldsymbol{E})$ and get the average of variance on the sampled set, which can reflect the balance of text representation on expressing the semantics of units. Table~\ref{var} shows that BERM has a smaller variance (semantic unit balance of text representation) and is more accurate in identifying the essential matching unit ( essential matching unit extractability of matching
representation) than baselines, which indicates the effectiveness of \textbf{R1} and \textbf{R2}.

\begin{table}[t]
\centering
\setlength\tabcolsep{3.5pt}
\scalebox{0.85}{
\begin{tabular}{lllllll}
\toprule
             & \multicolumn{2}{c}{DPR}                       & \multicolumn{2}{c}{KD}                        & \multicolumn{2}{c}{ANCE}                      \\ 
Axis
     & \multicolumn{1}{c}{X} & \multicolumn{1}{c}{Y} & \multicolumn{1}{c}{X} & \multicolumn{1}{c}{Y} & \multicolumn{1}{c}{X} & \multicolumn{1}{c}{Y} \\ \hline
Baseline     & 8.1                  & 6.5                  & 19.5                 & 18.5                 & 27.2                 & 25.9                 \\
BERM & 248.5                & 409.5                & 252.3                & 410.9                & 260.4                & 425.4  \\ \toprule           
\end{tabular}}
\caption{Dispersion of T-SNE result of representations of units in a passage (measured by the variance of the coordinates on x-axis and y-axis).}
\label{var-unit}
\end{table}
\textbf{Relationship Between Units.} Table~\ref{var-unit} shows that our method makes units in a passage more dispersed (tend to be orthogonal), which is more conducive to determining the unit that matches the query and masking the signals of other units. Our method makes the representation of the passage more suitable for matching, which is the domain-invariant feature for generalization.

\section{Conclusion}
In this paper, we propose an effective method called BERM to improve the generalization ability of dense retrieval without target domain data and additional modules. The basic idea of BERM is learning the domain-invariant feature, that is, matching signal. To achieve it, we introduce a novel concept of dense retrieval to represent the matching information between two texts, the matching representation. Further, we propose two requirements for matching and text representations as the constraint in the training of dense retrieval to enhance the ability to extract essential matching information from the passage according to different queries under the premise of balanced expression of the text. The two requirements unlock the ability of dense retrieval to capture matching signal without additional interaction. Experimental results show that BERM is a flexible method that can be combined with different dense retrieval training methods without inference overhead to improve the out-of-domain generalization ability. In domain adaptation setting, our method is also effective and performs better than baselines.

\section*{Limitations}
In this paper, we propose a novel concept of dense retrieval, the matching representation. Based on this, we introduce a novel generalizable dense retrieval training method via training the balanced and extractable representation for matching (BERM). Despite the strong performance of our method in improving the generalization ability of dense retrieval models, more theoretical proof needs to be researched to gain the deeper understanding of generalization improvement. Especially for matching representation, more theoretical analysis and implementation will be discussed in future work. We believe that the deeper study of matching representation will promote the development of dense retrieval, because it not only alleviates the problem that query and passage cannot interact in depth during training, but also describes the essence of retrieval task.

\section*{Ethics Statement}
Our work innovatively proposes the concept of matching representation in dense retrieval and designs a generalization improvement strategy that can be flexibly combined with different dense retrieval training methods. Our work has important implications for improving the performance of neural information retrieval models. We declare that our work complies with the ACL Ethics Policy.\footnote{\url{https://www.aclweb.org/portal/content/acl-code-ethics}}

\section*{Acknowledgements}
This work was supported by the National Key R\&D Program of China (2022YFB3103700, 2022YFB3103704), the National Natural Science Foundation of China (NSFC) under Grants No. 62276248, and the Youth Innovation Promotion Association CAS under Grants No. 2023111.

\normalem
\bibliography{anthology,my}
\bibliographystyle{acl_natbib}

\newpage

\appendix

\section{Datasets}
\label{Datasets}
In our experiment, source domain datsaet used as training data is MS-MARCO and target domain datasets used as testing data are collected from BEIR~\cite{beir}, which is a a heterogeneous benchmark to evaluate the generalization ability of retrieval models. Detials of the datasets are shown in Table~\ref{beir data}. In addition, we also introduce OAG-QA~\cite{oag-qa}, which is a fine-grained question-answering retrieval dataset consisting of different topics. We select datasets of different topics from 20 disciplines as the testing data to evaluate the generalization ability to different topics with different word distribution. Details of OAG-QA are shown in Table~\ref{oag data}.

\begin{table*}[t]
\renewcommand\arraystretch{1.25}
\centering
\scalebox{0.75}{
\begin{tabular}{lclcccc}
\toprule
\multirow{2}{*}{Task}          & \multirow{2}{*}{Domain} & \multirow{2}{*}{Dataset} & \multicolumn{2}{c}{Test} & \multicolumn{2}{c}{Avg. Word Lengths} \\ \cline{4-7} 
                               &                         &                          & \#Query    & \#Corpus      & Query            & Document           \\ \hline
Passage-Retrieval              & Misc.                   & MS-MARCO~\cite{msmarco}                 & 6,980     & 8,841,823    & 5.96             & 55.98              \\ \hline \hline
Bio-Medical                    & Bio-Medical             & TREC-COVID~\cite{treccovid}               & 50        & 171,332      & 10.60            & 160.77             \\
Information Retrieval          & Bio-Medical             & NFCorpus~\cite{nfcorpus}                 & 323       & 3,633        & 3.30             & 232.26             \\ \hline
Open-domain                    & Wikipedia               & NQ~\cite{nq}                       & 3,452     & 2,681,468    & 9.16             & 78.88              \\
Question                       & Wikipedia               & HotpotQA~\cite{hotpotqa}                 & 7,405     & 5,233,329    & 17.61            & 46.30              \\
Answering                      & Finance                 & FiQA-2018~\cite{fiqa}                & 648       & 57,638       & 10.77            & 132.32             \\ \hline
Argument                       & Misc.                   & ArguAna~\cite{arguana}                  & 1,406     & 8,674        & 192.98           & 166.80             \\
Retrieval                      & Misc.                   & Touché-2020~\cite{touche}              & 49        & 382,545      & 6.55             & 292.37             \\ \hline
Duplicate-Question             & StackEx.                & CQADupStack~\cite{cqadupstack}              & 13,145    & 457,199      & 8.59             & 129.09             \\
Retrieval                      & Quora                   & Quora~\cite{beir}                    & 10,000    & 522,931      & 9.53             & 11.44              \\ \hline
Entity-Retrieval               & Wikipedia               & DBPedia~\cite{dbpedia}                  & 400       & 4,635,922    & 5.39             & 49.68              \\
Citation-Prediction            & Scientific              & SCIDOCS~\cite{scidocs}                  & 1,000     & 25,657       & 9.38             & 176.19             \\ \hline
\multirow{3}{*}{Fact Checking} & Wikipedia               & FEVER~\cite{fever}                    & 6,666     & 5,416,568    & 8.13             & 84.76              \\
                               & Wikipedia               & Climate-FEVER~\cite{climatefever}            & 1,535     & 5,416,568    & 20.13            & 84.76              \\
                               & Scientific              & SciFact~\cite{scifact}                  & 300       & 5,183        & 12.37            & 213.63             \\ 
\toprule
\end{tabular}
}
\caption{Details of the datasets in BEIR, the table is collected from~\cite{beir}.}
\label{beir data}
\end{table*}

\begin{table}[t]
\setlength\tabcolsep{2pt}
\centering
\scalebox{0.72}{
\begin{tabular}{llll}
\toprule
Discipline & Topic                    & \#Query & \#Corpus \\ \hline
Geometry    & Geometry                 & 230                   & 10,000              \\
Statistics  & Mathematical Statistics  & 144                   & 10,000              \\
Algebra     & Polynomial               & 280                  & 10,000              \\
Calculus    & Calculus                 & 242                   & 10,000             \\
Number theory  & Number theory            & 274               & 10,000              \\
Linear algebra  & Matrix                   & 130                  & 10,000              \\
Astrophysics    & Black hole               & 160                  & 10,000             \\
Physics       & Classical mechanics      & 115                  & 10,000              \\
Chemistry     & Physical chemistry       & 190                   & 10,000              \\
Biochemistry  & Biochemistry             & 129                   & 10,000             \\
Health Care   & Health care              & 288                   & 10,000             \\
Natural Science & Evolutionary biology     & 471                  & 10,000             \\
Psycology       & Cognitive neuroscience   & 348                   & 10,000              \\
Algorithm      & Algorithm                & 386                   & 10,000             \\
Neural Network  & Neural network           & 590                  & 10,000             \\
Data Mining     & Data mining              & 131                 & 10,000             \\
Computer Graphics   & Computer graphics images & 68                  & 10,000             \\
Deep Learning       & Optimization             & 238                  & 10,000             \\
Machine Learning    & Linear regression        & 244                  & 10,000             \\
Economics           & Economics                & 238                  & 10,000    \\ \toprule        
\end{tabular}
}
  \caption{Details of QAG-QA.}
\label{oag data}
\end{table}

\section{Baselines}
\label{Baseline}
We introduce the baselines in the main experiment and the domain adaptation experiment respectively. 
\subsection{Baselines for Main Experiment}
In the main experiment, our method is combined with different mainstream dense retrieval training methods to improve its generalization. We consider three training methods including vanilla (DPR~\cite{dpr}), knowledge distillation (KD) and hard negatives mining (ANCE~\cite{ance}).
\begin{itemize}
    \item \textbf{DPR} trains the dense retrieval model via in-batch negative sampling. Different from \cite{dpr}, we train DPR on MS-MARCO to achieve a fair comparison. 
    \item \textbf{KD} trains the dense retrieval model under the guidance of the soft labels provided by the teacher model. In the experiment, we use a cross-encoder model trained on MS-MARCO as the teacher model.
    \item \textbf{ANCE} trains the dense retrieval model with hard negatives updated in parallel as described in~\cite{ance}. 
\end{itemize}
\subsection{Baselines for Domain Adaptation}
\begin{itemize}
    \item \textbf{MoDIR} uses the data from source and target domains for adversarial training to perform unsupervised domain adaptation. 
    \item \textbf{Contriever} performs unsupervised pre-training on Wikipedia and CC-Net~\cite{ccnet}.
    \item \textbf{GenQ} uses T5~\cite{t5} generates 5 queries for each passage in target domain and fine-tunes TAS-B~\cite{tas-b} on this data.
    \item \textbf{GPL} improves the domain adaptation performance based on GenQ. In addition to generated queries, GPL uses cross-encoder to provide the pseudo-label. GPL fine-tunes multiple backbones on the generated queries and pseudo-labels and we report the best performance that is fine-tuned on TAS-B.
    \item \textbf{COCO-DR} performs unsupervised pre-training on target domain and introduces distributional robust optimization.
\end{itemize}

\begin{figure*}[t]
    \centering
            \subfigure[Text Rep.]{
	\includegraphics[width=1.47in]{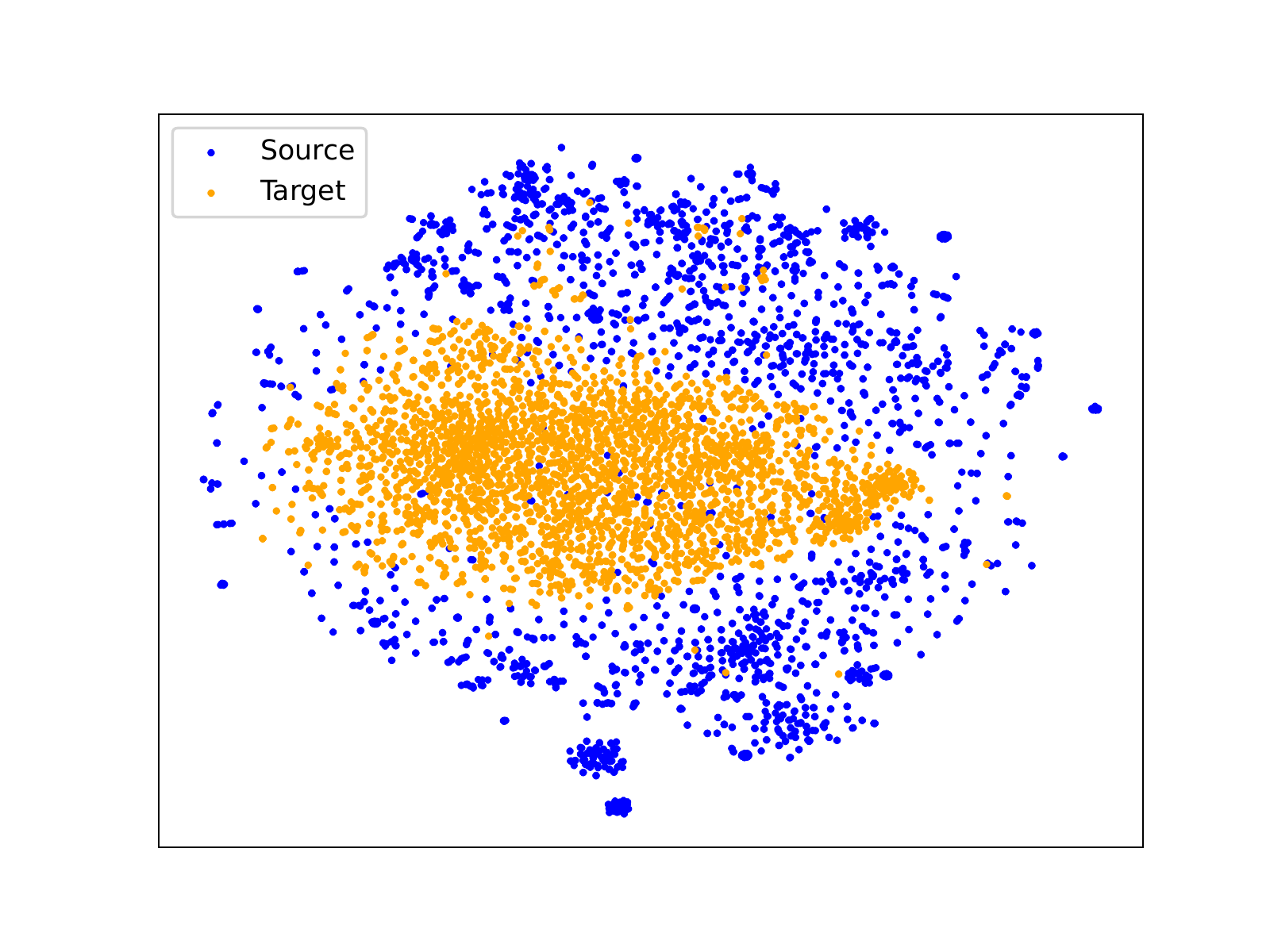}
	
    }
    \subfigure[Text Rep. (ours)]{
        \includegraphics[width=1.47in]{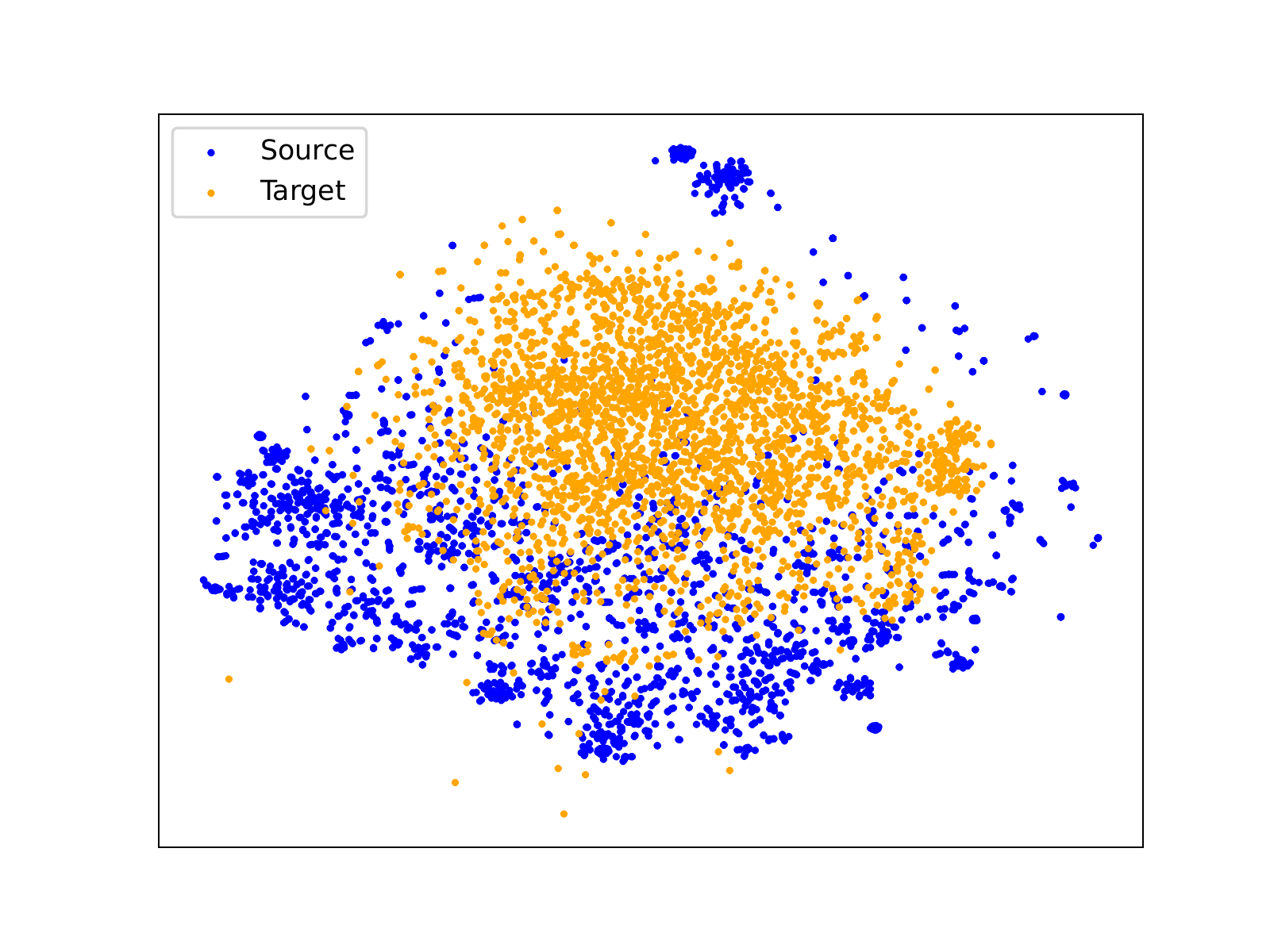}
    
    }
    \subfigure[Matching Rep.]{
        \includegraphics[width=1.47in]{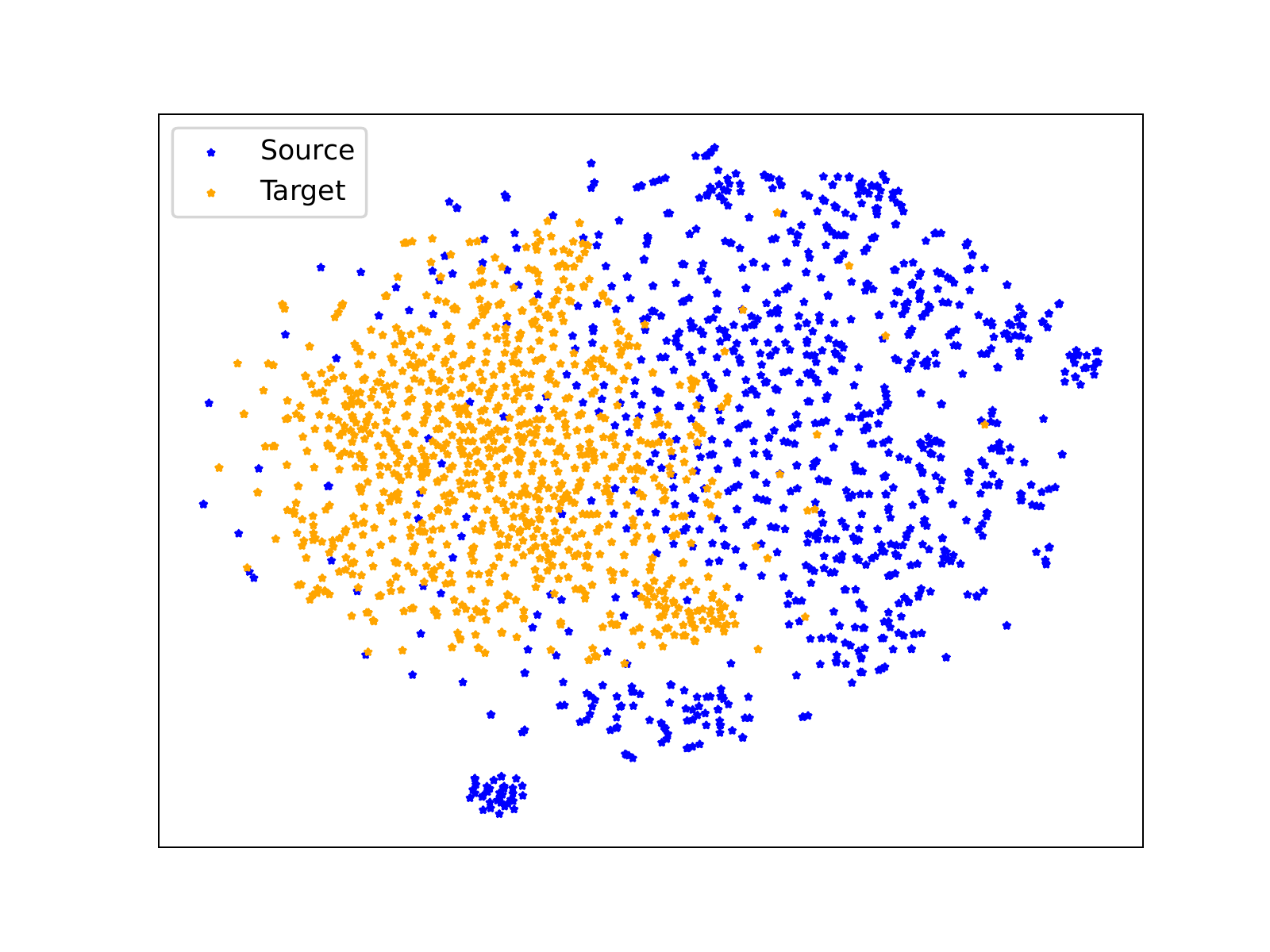}
    
    }
    \subfigure[Matching Rep. (ours)]{
        \includegraphics[width=1.47in]{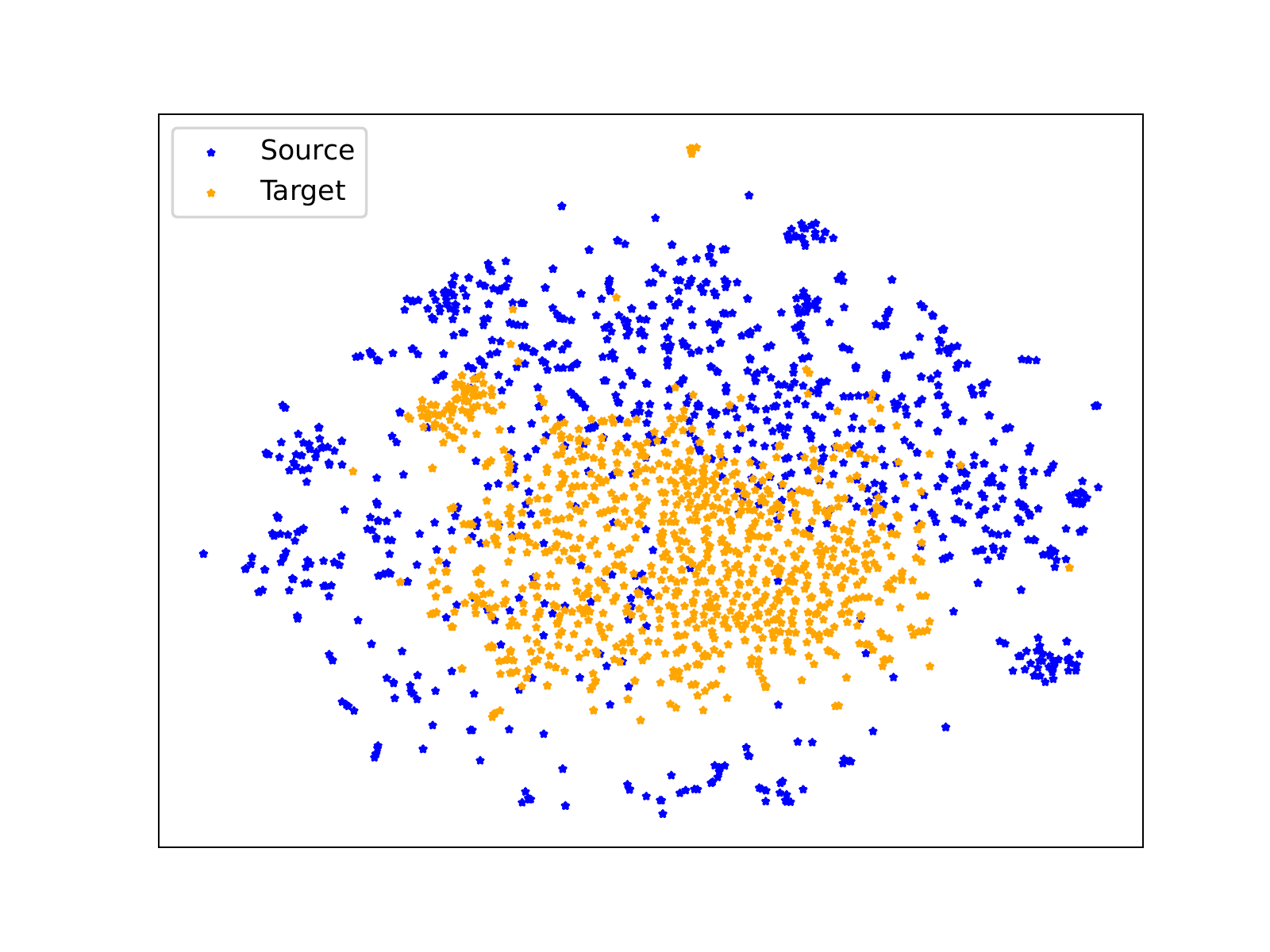}
    
    }
        \subfigure[Text Rep.]{
	\includegraphics[width=1.47in]{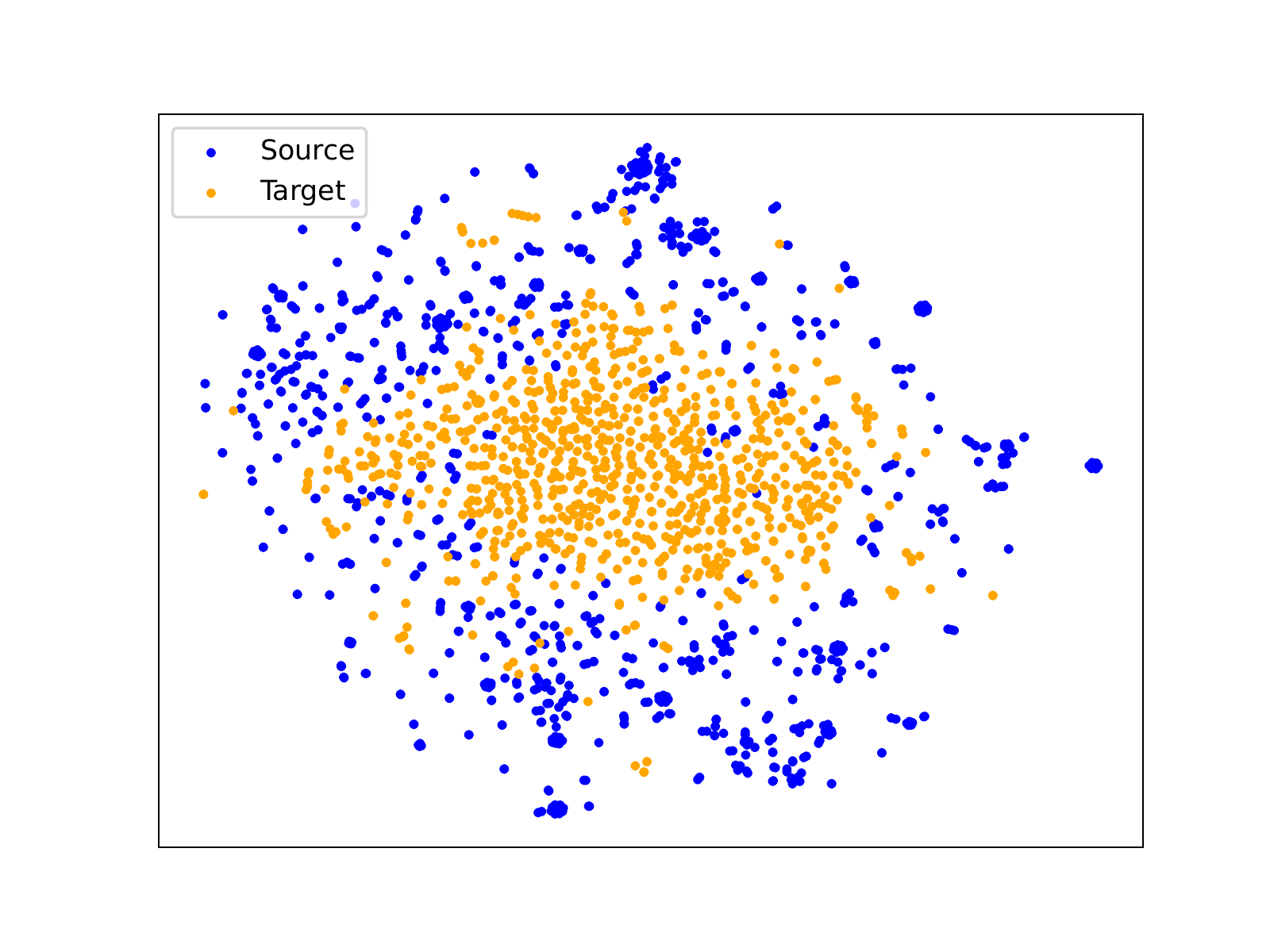}
    }
    \subfigure[Text Rep. (ours)]{
        \includegraphics[width=1.47in]{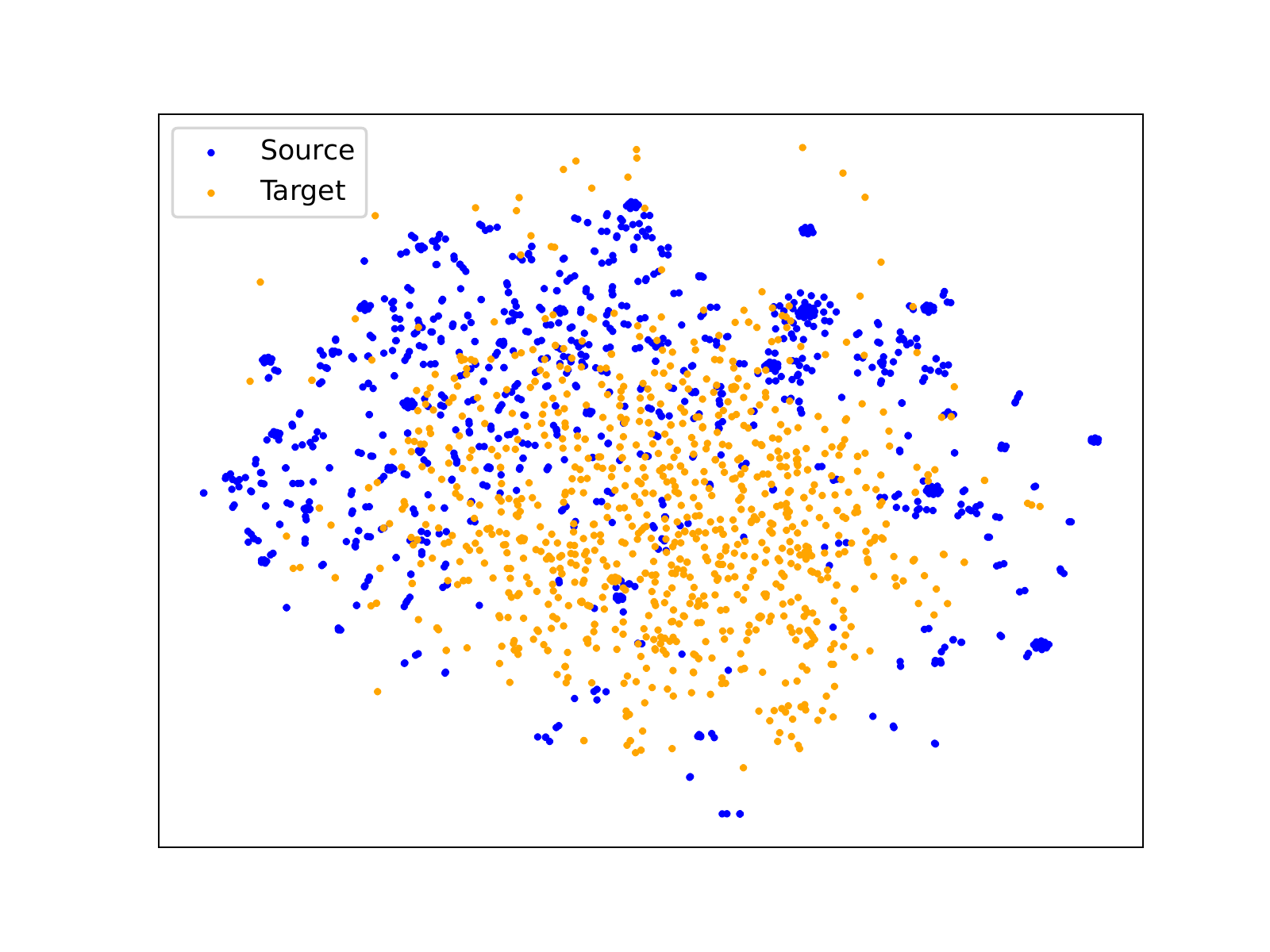}
    }
    \subfigure[Matching Rep.]{
        \includegraphics[width=1.47in]{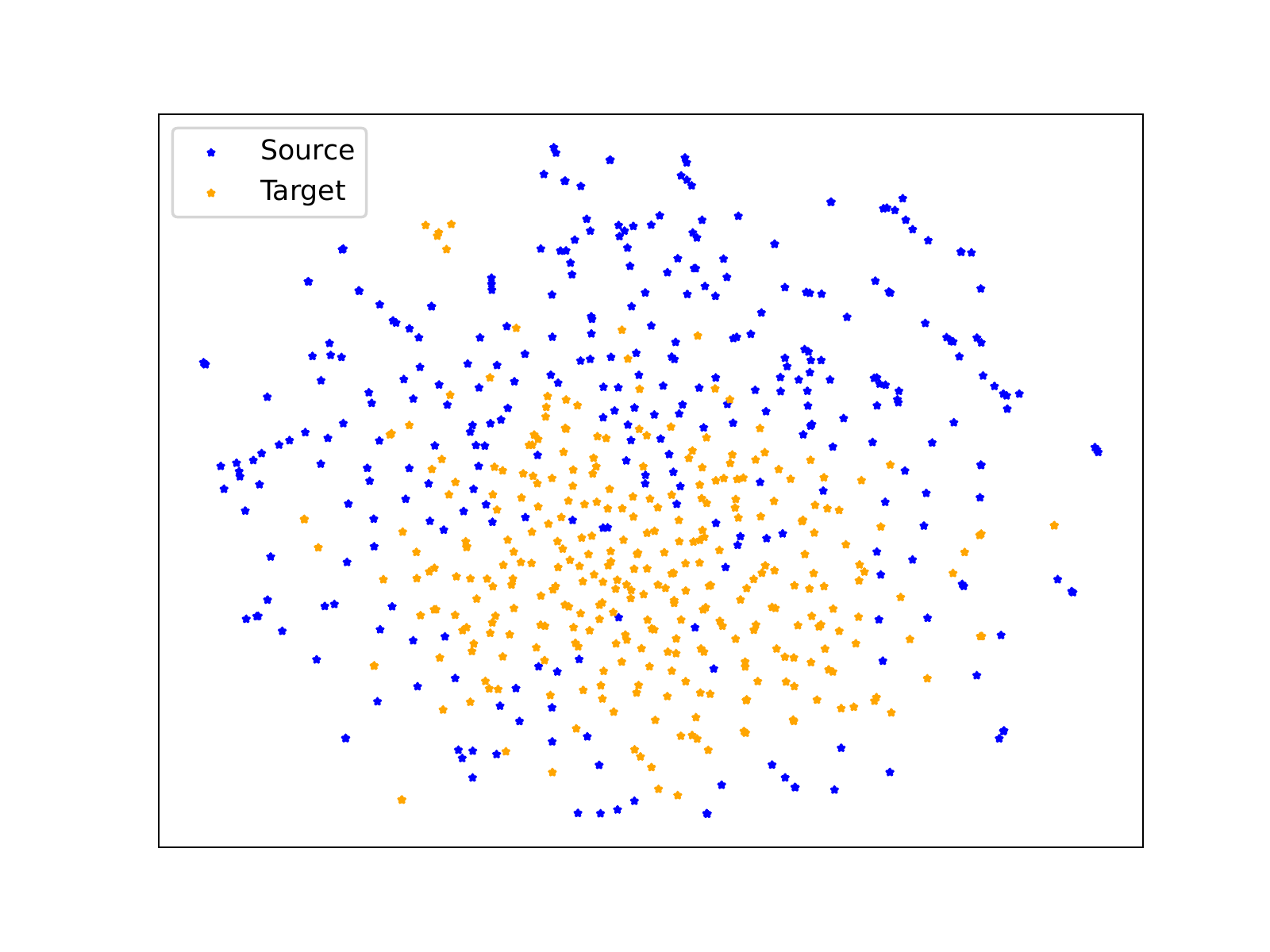}
    }
    \subfigure[Matching Rep. (ours)]{
        \includegraphics[width=1.47in]{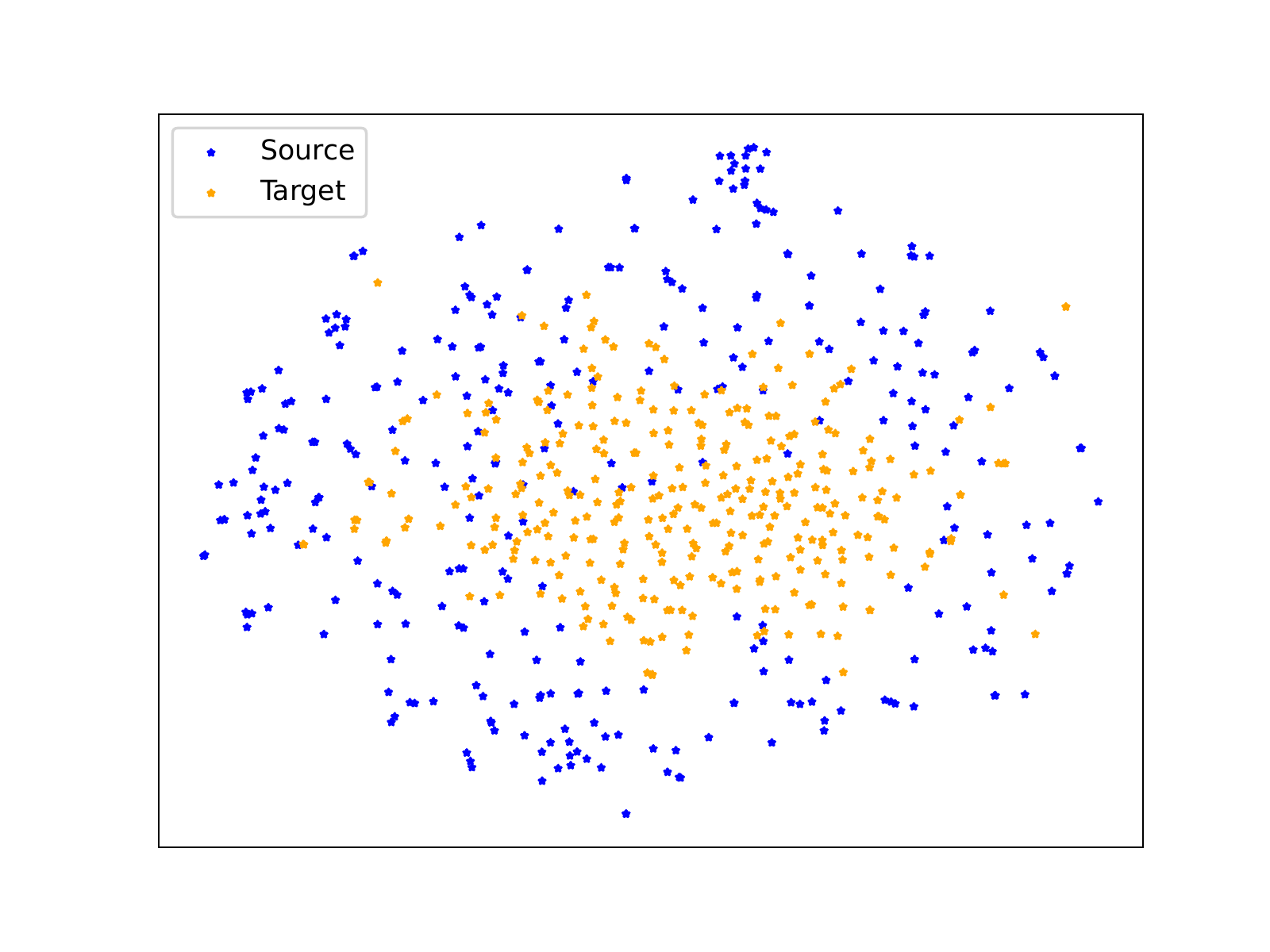}
    }

    \subfigure[Text Rep.]{
	\includegraphics[width=1.47in]{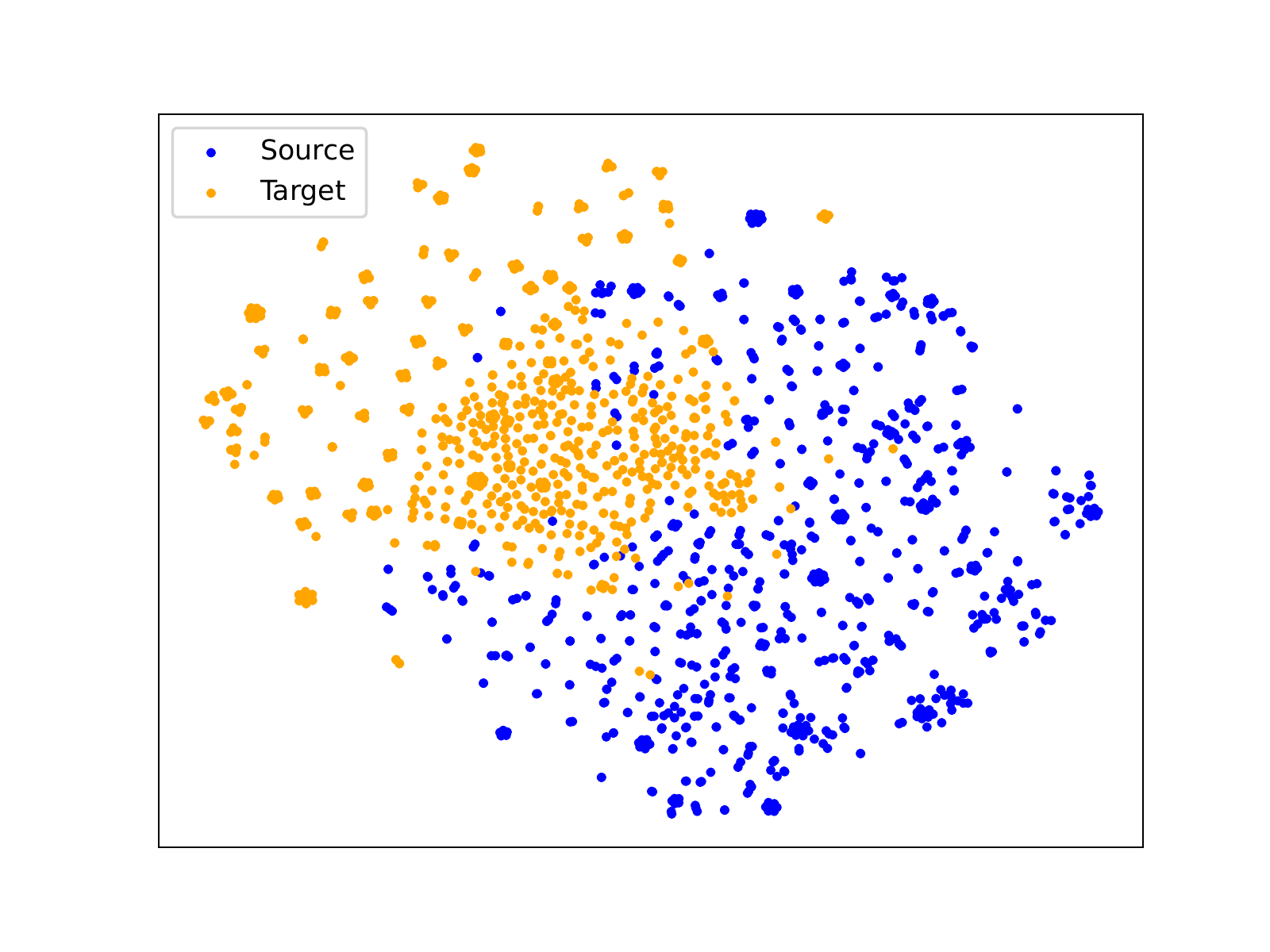}
	
    }
    \subfigure[Text Rep. (ours)]{
        \includegraphics[width=1.47in]{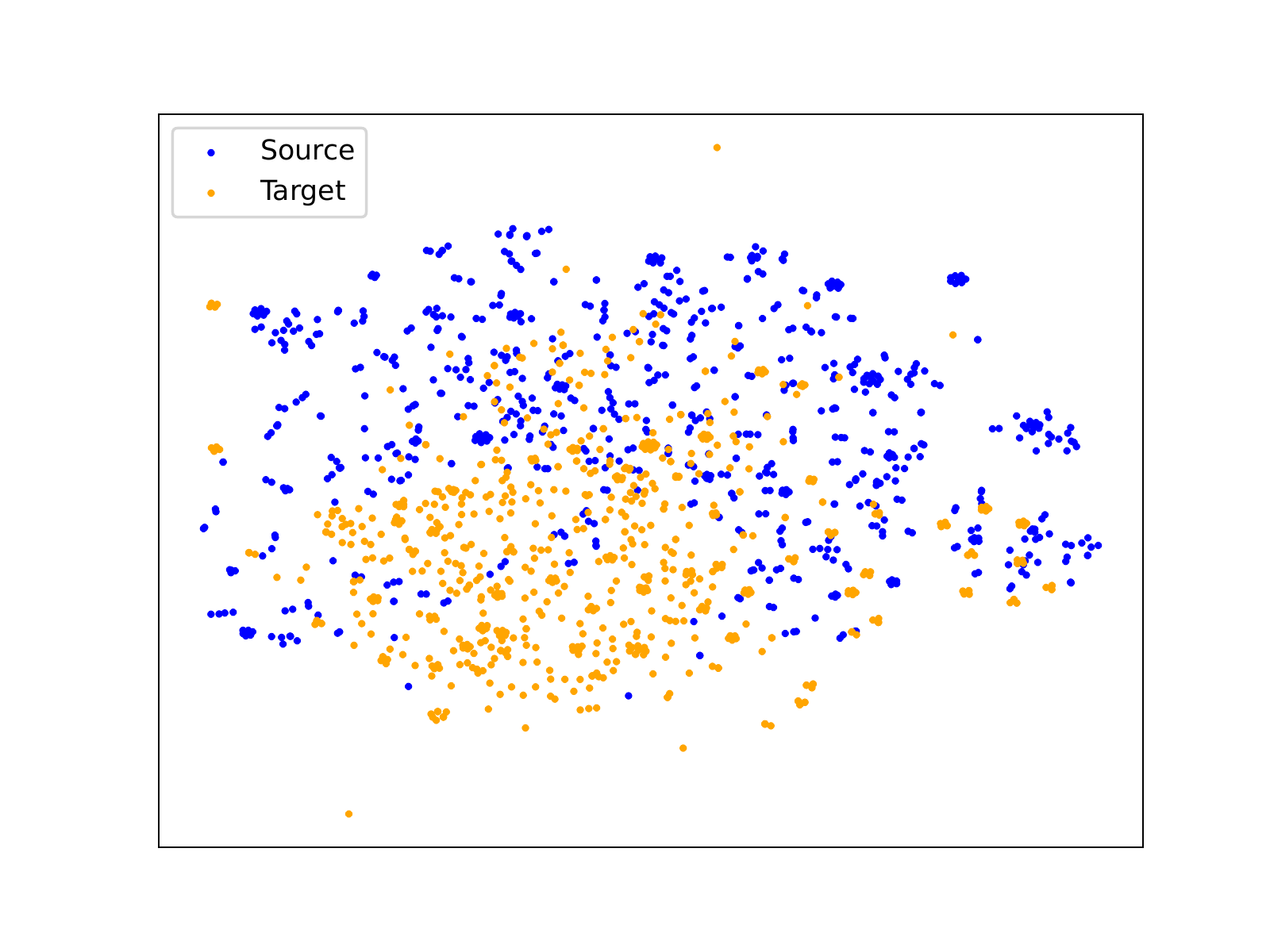}
    
    }
    \subfigure[Matching Rep.]{
        \includegraphics[width=1.47in]{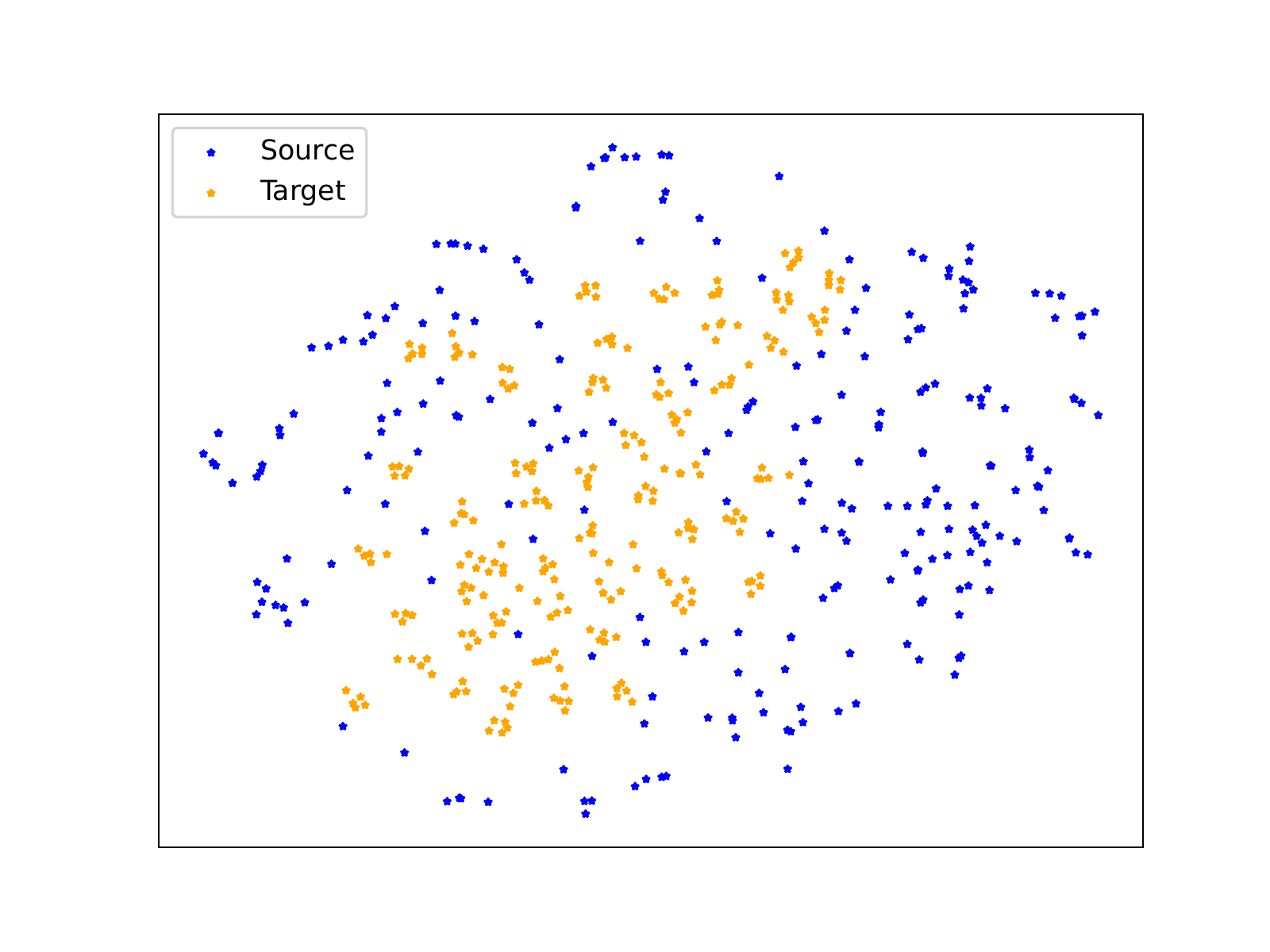}
    
    }
    \subfigure[Matching Rep. (ours)]{
        \includegraphics[width=1.47in]{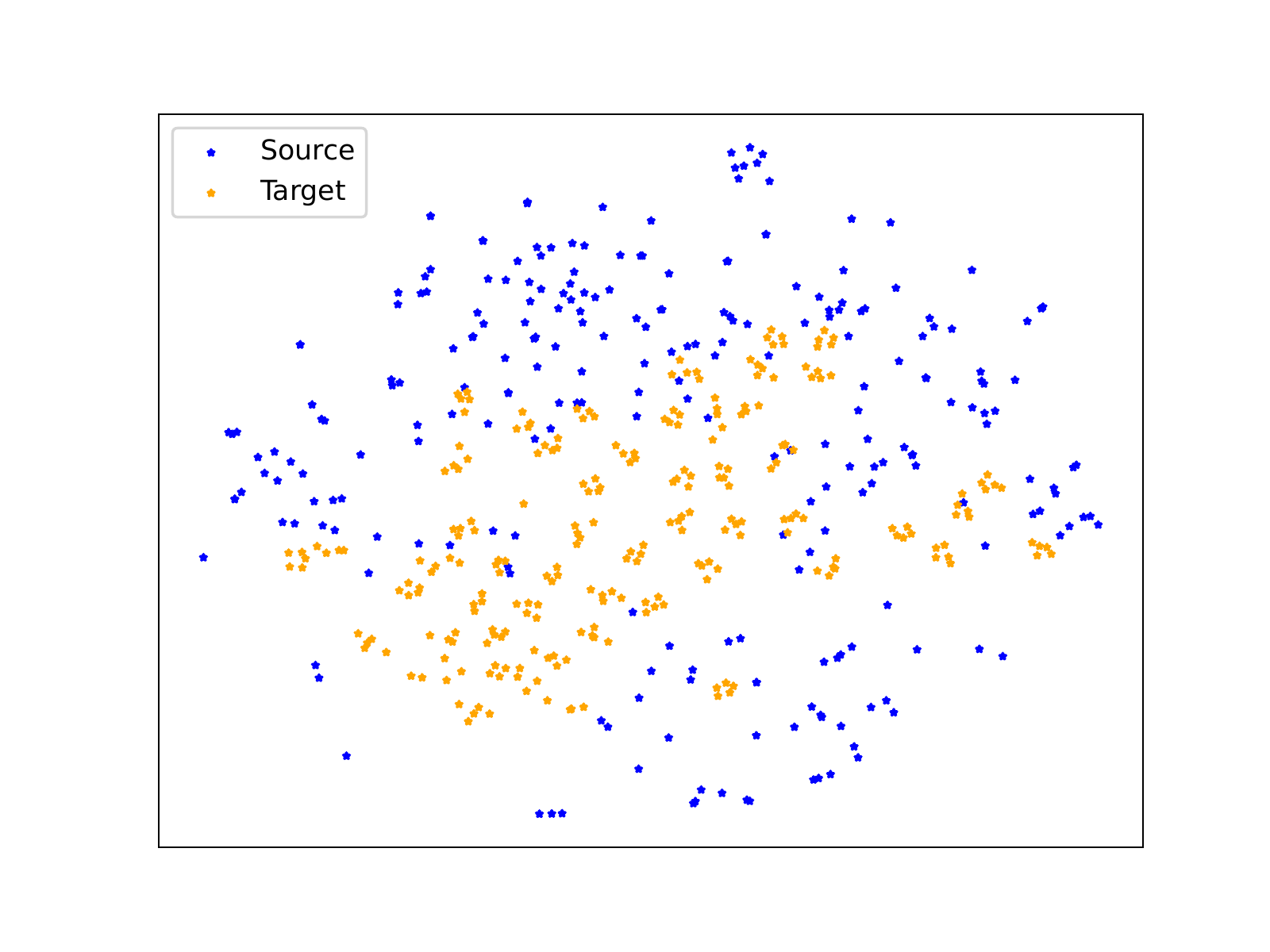}
    
    }

    \subfigure[Text Rep.]{
	\includegraphics[width=1.47in]{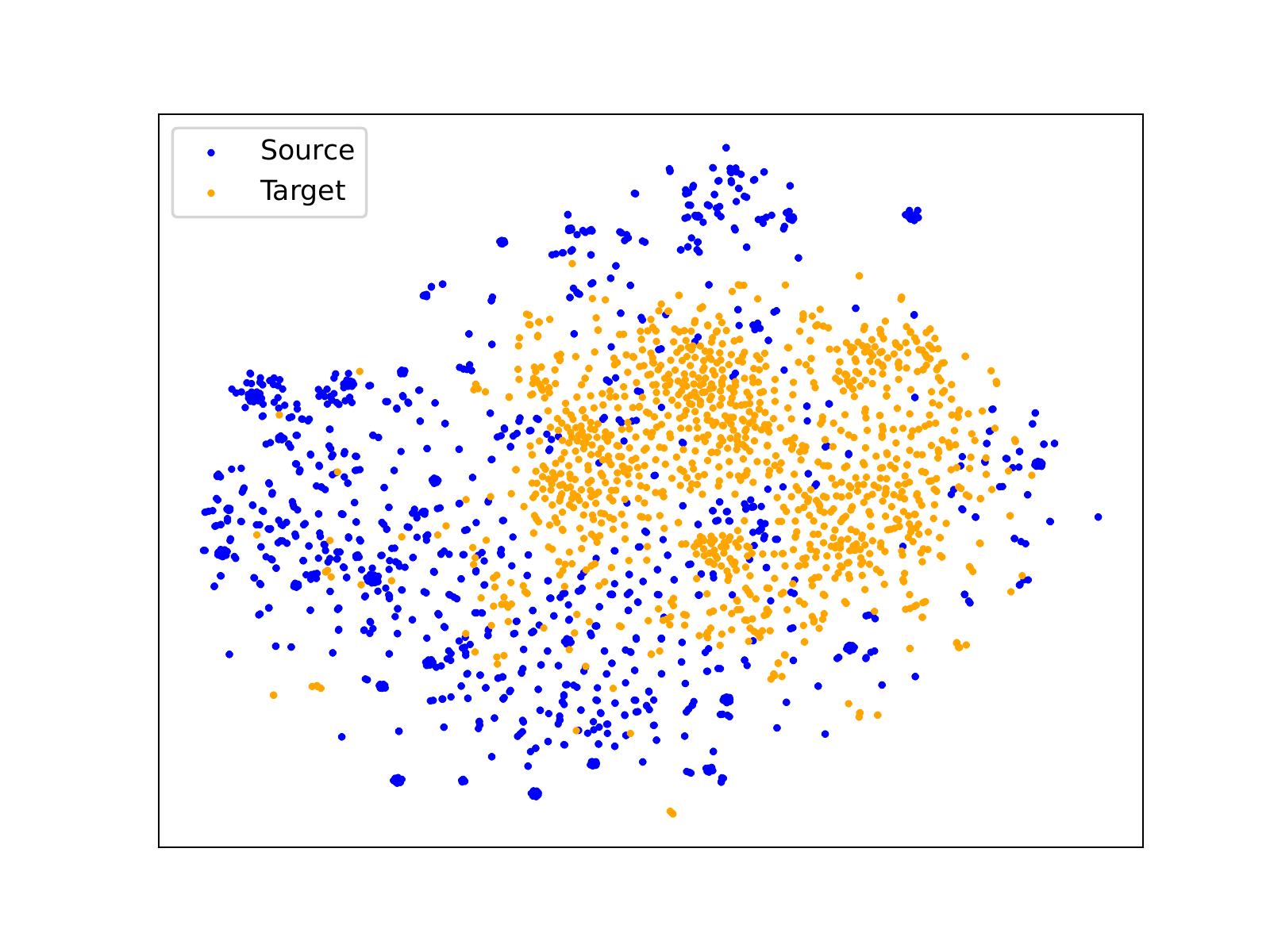}
	
    }
    \subfigure[Text Rep. (ours)]{
        \includegraphics[width=1.47in]{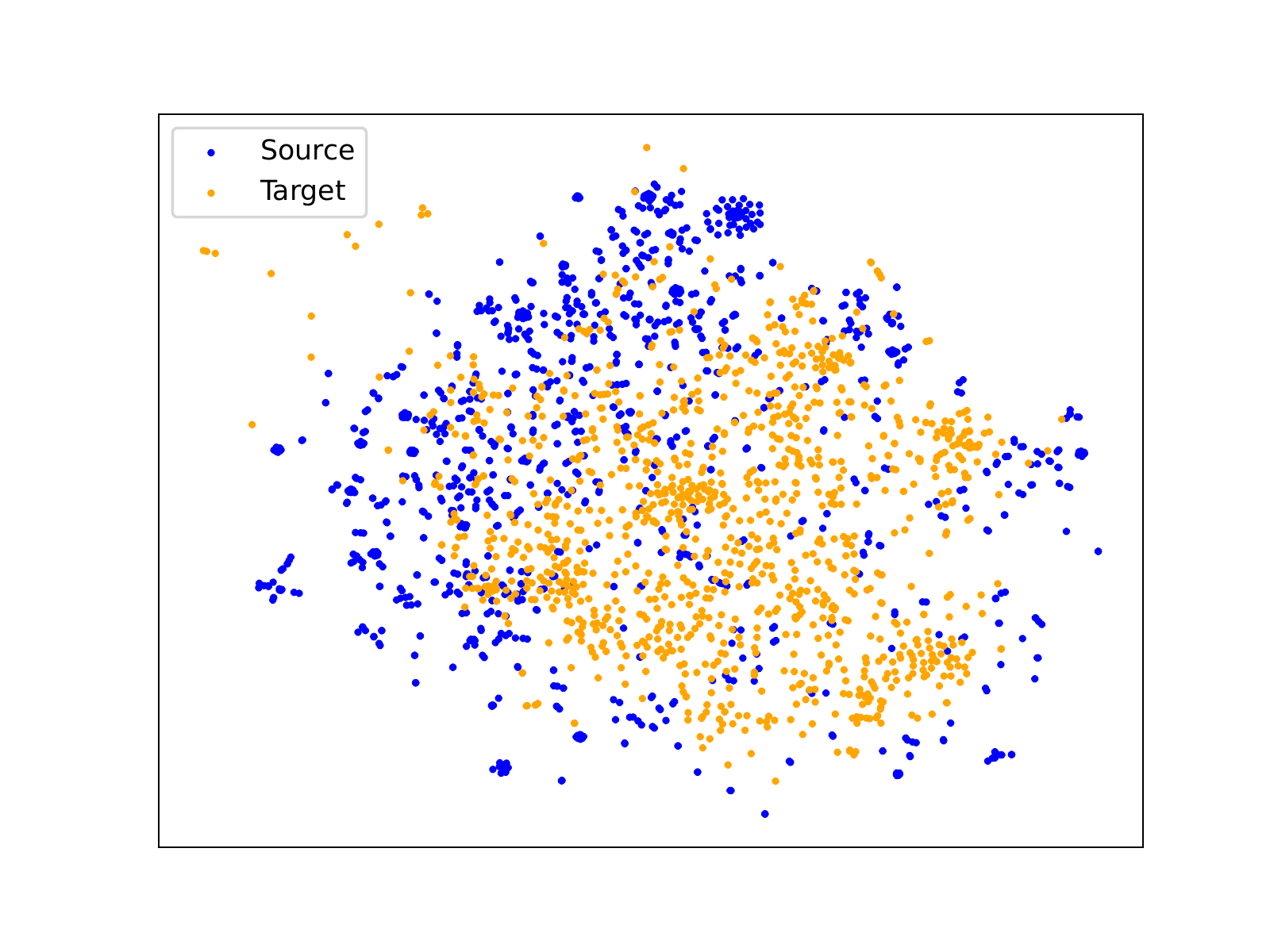}
   
    }
    \subfigure[Matching Rep.]{
        \includegraphics[width=1.47in]{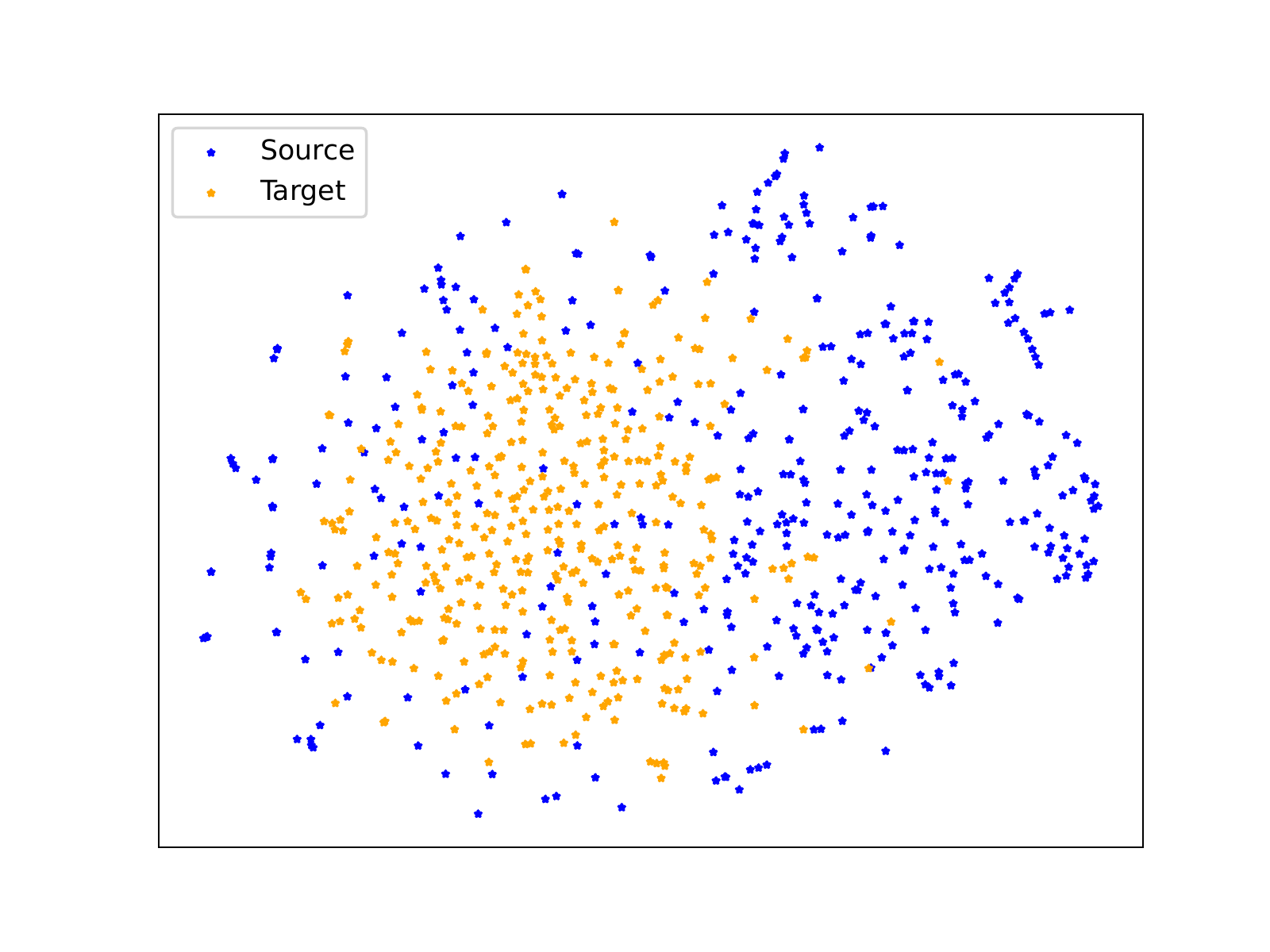}
   
    }
    \subfigure[Matching Rep. (ours)]{
        \includegraphics[width=1.47in]{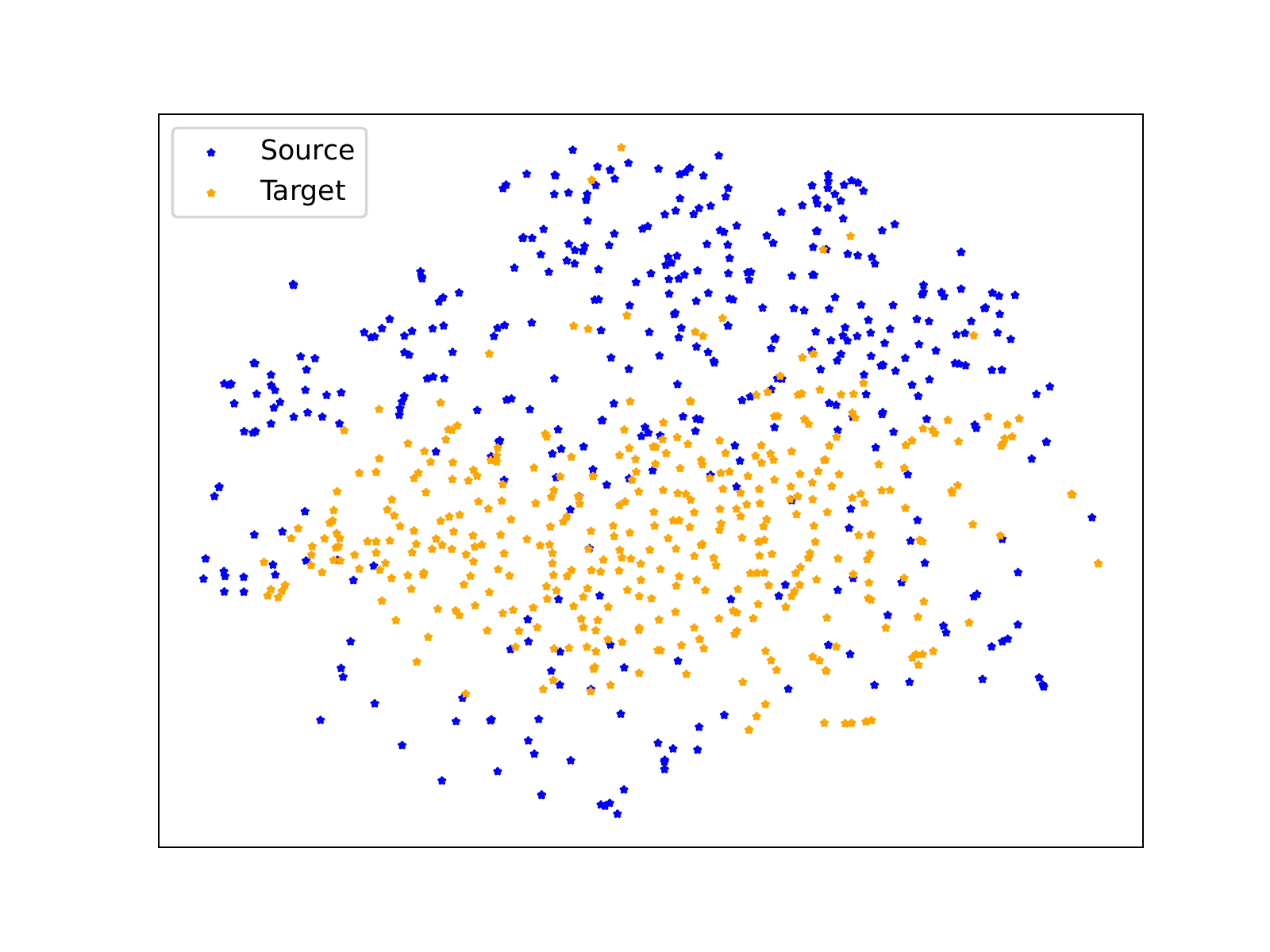}
    
    }
\caption{T-SNE of the text and matching representations for source and target domains.}
\label{domain-invariant more}
\end{figure*}
\section{Domain-invariant Representation} \label{domain-figure}

Visualized results of T-SNE of representations of source and target (SCIDOCS, TREC-COVID, NFCorpus and DBpedia) domains encoded by DPR and DPR+BERM respectively are shown in Figure~\ref{domain-invariant more}.

\end{document}